# A Real-Time Closed-Form Model for Nonlinearity Modeling in Ultra-Wide-Band Optical Fiber Links Accounting for Inter-channel Stimulated Raman Scattering and Co-Propagating Raman Amplification

**Mahdi Ranjbar Zefreh, Pierluigi Poggiolini**

*Politecnico di Torino, DET, Torino, Italy, mahdi.ranjbarzefreh@polito.it*

**Abstract:** In this paper, we present a novel closed-form model (CFM) for accurate and fast evaluation of nonlinear interference in modern ultrawideband coherent optical fiber communication systems. Starting from the Gaussian noise model (GN model), using reasonable approximations and machine-learning optimized improvements, we achieve an accurate CFM capable of handling ultrawide band (C+L or wider) optical fiber systems in the presence of Inter-channel Stimulated Raman Scattering (ISRS) and forward-pumped Raman amplification.

## 1- Introduction

Physical-layer-aware control and optimization of ultra-high-capacity optical networks is becoming an increasingly important aspect of networking, as throughput demand and loads increase. A necessary pre-requisite is the availability of accurate analytical modeling of fiber non-linear effects (or NLI, Non-Linear-Interference).

Several NLI models have been proposed over the years, such as 'time-domain' [1], [2], GN [3], EGN [4] , [5] , others such as [6]-[9], and various precursors of the all of them (see for instance refs. in [10]). These NLI models, however, either contain integrals that make them unsuitable for real-time use, or otherwise assume too idealized system set-ups. The challenge is to derive approximate closed-form formulas, thus enabling real-time computation, that both preserve accuracy and are general enough to model highly diverse actual deployed systems.

In the GN/EGN model class, a rather general closed-form set of formulas (or closed-form model, CFM) has been available for several years (Eqs. (41)-(43) in [3]). These formulas, that we call 'CFM0', are a closed-form approximation of the incoherent GN-model (or iGN model [3]).  They already allow to deal with systems with arbitrary WDM combs and non-identical spans and amplifiers. However, they do not support, among other things, dispersion slope and frequency-dependent loss, all-important features to enable the real-time modeling of actually-deployed realistic systems and networks.

We upgraded CFM0 to include such missing features, following the approach proposed in [11], similar (though not identical) to [12]. We call these new formulas CFM1. We first tested CFM1 over the C-band, to make sure that it performed well in that context, neglecting ISRS. We did the testing over a very large number (8500) of highly-randomized C-band WDM systems, comparing the results of CFM1 with a highly accurate version of the EGN model (numerically integrated).

CFM1 performed acceptably well, but it showed an average tendency towards overestimating NLI. This could be expected since CFM1 is an approximation of the GN-model, whose known behavior is to somewhat overestimate NLI [3]. In addition to such pessimistic bias, we also observed a non-negligible variance of the error.

To improve the accuracy of CFM1 vs. the EGN benchmark, we leveraged the 8500 systems test-set to find a simple correction law which contained both physical system parameters and best-fitted coefficients, with the goal of turning CFM1 from a GN-model emulating CFM into an accurate EGN-emulating CFM. This approach, that can be viewed as machine-learning over a big-data set, proved effective [13]-[16]. The new set of formulas, called CFM2, had a much lower NLI estimation error.

Still, we could observe rare elevated error outliers in NLI estimation. We managed to remove them by improving CFM2 using a further analytical contribution, which was derived as shown in [17]. This further contribution was designed to account for coherence effects in the accumulation of NLI noise. With the addition of such term, the new closed-form model (CFM3) performed better on all accounts and, in particular, removed all outliers, drastically curtailing the peak error A final refinement consisted in accounting for the effect of channel roll-off, resulting in CFM4, which was also tested over the 8500 systems test-set [16].

CFM4 provides a real-time very effective and very accurate tool for NLI computation, in arbitrary C-band systems, with the only limitation of fiber dispersion not going below about 2 ps/(nm km). Under this value, some of the approximations used to derive the CFMs break down. Separate work is currently ongoing to extend CFM4 towards lower and near-zero dispersion [18],[19] but we consider this outside the scope of this paper.

The aim of this work is instead to extend CFM4 so that it can deal with C+L (or even broader band) systems, while retaining its excellent flexibility, accuracy, and real-time computation capability, in scenarios where dispersion does not go below 2 ps/(nm km).

To do so we re-start from the premises of [11] and propose a detailed re-derivation of CFM4. However, we introduce some substantial changes so that a more powerful set of formulas is achieved. The new CFM, called CFM5, is capable of modeling NLI in any arbitrary ultra-broadband WDM system (C+L or wider) accounting for ISRS and forward-pumped Raman amplification, with high accuracy and real-time computation speed.

The current main remaining limitations of CFM5 are the requirement of dispersion not being lower than 2 ps/(nm km) and the need to separately compute the SRS-induced power evolution for each channel along the link, which is comparatively much slower than CFM5 itself (seconds vs. tens of milliseconds).

Work is ongoing on cutting down SRS computation time, to try to achieve an all-encompassing, real-time general NLI model capable of handling arbitrary ultra-broadband systems. In this paper, we focus on CFM5 and leave the SRS computation speed-up part for a subsequent paper.

## 2- ISRS mathematical modeling

Stimulated Raman Scattering (SRS) is a well-known nonlinear effect which causes a power transfer from higher frequencies (shorter wavelengths) to lower frequencies (longer wavelengths). In a WDM frequency comb propagating in the fiber, containing $N$ channels with center frequencies $f_1 < f_2 < \cdots < f_N$, the propagation-distance dependent power of each channel can be modeled by a set of coupled differential equations as [20]:

$$\begin{cases} \frac{dP_1}{dz} = \{\sum_{i=2}^{N} C_R(f_i - f_1) \times P_i\} \times P_1 - 2 \times \alpha_1 \times P_1 \\ \vdots \\ \frac{dP_l}{dz} = \{-\sum_{i=1}^{(l-1)} \frac{f_l}{f_i} \times C_R(f_l - f_i) \times P_i\} \times P_l + \{\sum_{i=(l+1)}^{N} C_R(f_i - f_l) \times P_i\} \times P_l - 2 \times \alpha_l \times P_l \\ \vdots \\ \frac{dP_N}{dz} = \{-\sum_{i=1}^{(N-1)} \frac{f_N}{f_i} \times C_R(f_N - f_i) \times P_i\} \times P_N - 2 \times \alpha_N \times P_N \end{cases} \quad \text{eq. (1)}$$

Where eq. (1) is a set of $N$ coupled nonlinear differential equations. $z$ is the distance from the signal launch location in the fiber. $l$ can be in the range $2 \leq l \leq (N-1)$ and $P_j(0)$ is the power launched into the $j$'th channel at the start of the fiber span while $P_j = P_j(z)$ is the power of the $j$'th channel at the distance $z$ ($1 \leq j \leq N$). Also, $\alpha_j$ is the fiber loss parameter for the $j$'th channel in the absence of Raman and in general, can be different channel by channel. We may also denote $\alpha_j$ by $\alpha(f_j)$ in this paper so $\alpha_j = \alpha(f_j)$. $C_R(u)$ is an odd function with respect to its frequency variable $u$ which represents the gain profile of the ISRS effect and it depends on the fiber physical specifications. For $u > 0$, $C_R(u) \geq 0$ and for $u < 0$, $C_R(u) = -C_R(-u)$.

It is worth noting that in the absence of the ISRS effect, $C_R(u) = 0 \ \forall u$, and eq. (1) has the obvious analytical solution: $P_k(z) = P_k(0) \times e^{-2 \times \alpha_k \times z}$ for $1 \leq k \leq N$.

In general, there is not an analytical solution for eq. (1), which then must be solved numerically. Under certain specific conditions, however, an approximate analytical solution can be found, which we discuss it in the next section.

## 3- Analytical Solution of ISRS differential equations under specific conditions

As we mentioned in the previous section, eq. (1) has an obvious analytical solution in the absence of SRS ($C_R(u) = 0 \ \forall u$) which is $P_k(z) = P_k(0) \times e^{-2 \times \alpha_k \times z}$ for $1 \leq k \leq N$. When SRS is present ($C_R(u) \neq 0$) there is an analytical solution provided that the 7 specific conditions hold [21]. Before listing these conditions and showing the related closed-form solution, we would like to point out that in actual systems the 7 conditions may be only marginally met, or not met.

However, this is not important since we are discussing here this closed-form solution because we then use it to draw inspiration to write a more general and effective formula.

1. If $C_R(u)$ is considered as a triangular shape function as:

$$C_R(u) = \begin{cases} \dfrac{C_{R,max}}{\Delta f_{ISRS}} \times u & 0 \leq |u| \leq \Delta f_{ISRS} \\ 0 & |u| > \Delta f_{ISRS} \end{cases} \qquad eq.\ (2)$$

where in (2) $C_{R,max}$ and $\Delta f_{ISRS}$ are two positive constants called *ISRS gain profile maximum* and *ISRS bandwidth* respectively. Considering $C_R(u)$ as shown in eq. (2) is called the 'triangular approximation' of the Raman gain profile.

2. $(f_N - f_1) \leq \Delta f_{ISRS}$ which indicates the total WDM bandwidth should be less than the ISRS bandwidth.

3. $\alpha_1 = \alpha_2 = \cdots = \alpha_{(N-1)} = \alpha_N = \alpha_0$, i.e., identical loss is assumed for all frequencies (channels) in the absence of ISRS.

4. $(f_2 - f_1) = (f_3 - f_2) = \cdots = (f_{(j+1)} - f_j) = \cdots = (f_{(N-1)} - f_{(N-2)}) = (f_N - f_{(N-1)}) = \Delta f_{ch}$, that is all channels are equally spaced in the WDM comb.

5. Replacing terms $\dfrac{f_l}{f_i}$ and $\dfrac{f_N}{f_i}$ with 1 in eq. (1) ($\dfrac{f_l}{f_i} \cong 1$ and $\dfrac{f_N}{f_i} \cong 1\ \forall i, l$). This approximation has an interesting physical interpretation. From a quantum mechanics point of view, in the ISRS nonlinear process a photon with high frequency (high energy) is converted into a lower frequency (lower energy) photon. Therefore, during this conversion, there is some loss of energy. By assuming $\dfrac{f_l}{f_i} \cong 1$ and $\dfrac{f_N}{f_i} \cong 1\ \forall i, l$, we basically neglect the photon conversion loss.

With the above five conditions, eq (1) can be solved analytically as [21]:

$$P_j(z) = P_j(0) \times e^{-2\alpha_0 z} \times \dfrac{P_{tot} \times e^{P_{tot} \times \frac{C_{R,max}}{\Delta f_{ISRS}} \times (f_N - f_j) \times \frac{1-\exp(-2\alpha_0 z)}{2\alpha_0}}}{\sum_{i=1}^{N} P_i(0) \times e^{P_{tot} \times \frac{C_{R,max}}{\Delta f_{ISRS}} \times (f_N - f_i) \times \frac{1-\exp(-2\alpha_0 z)}{2\alpha_0}}} \qquad eq.\ (3)$$

where $P_{tot}$ the total input power for all WDM channels defined as:

$$P_{tot} \triangleq \sum_{k=1}^{N} P_k(0) \qquad eq.\ (4)$$

With one further assumption:

6- $P_1(0) = P_2(0) = \cdots = P_j(0) = \cdots = P_{N-1}(0) = P_N(0) = P_0$ ; that means equal power (uniform power) for all WDM channels. We have:

$$P_{tot} = \sum_{k=1}^{N} P_k(0) = NP_0 \qquad\qquad eq.~(5)$$

then eq. (3) simplifies to [21]:

$$P_j(z) = N \times P_0 \times e^{-2\alpha_0 z} \times \frac{\sinh\left(\frac{NP_0}{2} \times \frac{C_{R,max}}{\Delta f_{ISRS}} \times \Delta f_{ch} \times \frac{1-\exp(-2\alpha_0 z)}{2\alpha_0}\right)}{\sinh\left(\frac{N^2 P_0}{2} \times \frac{C_{R,max}}{\Delta f_{ISRS}} \times \Delta f_{ch} \times \frac{1-\exp(-2\alpha_0 z)}{2\alpha_0}\right)}$$

$$\times \left\{ e^{\frac{NP_0}{2} \times \frac{C_{R,max}}{\Delta f_{ISRS}} \times (f_N + f_1 - 2f_j) \times \frac{1-\exp(-2\alpha_0 z)}{2\alpha_0}} \right\} \qquad eq.~(6)$$

We finally make one more assumption as:

7- $N^2 P_0 \times \frac{C_{R,max}}{\Delta f_{ISRS}} \times \Delta f_{ch} \times \frac{1-\exp(-2\alpha_0 z)}{2\alpha_0} \ll 1, \forall z$, or equivalently as $\left\{ P_{tot} \times \frac{C_{R,max}}{\Delta f_{ISRS}} \times BW_{WDM} \times \frac{1-\exp(-2\alpha_0 z)}{2\alpha_0} \right\} \ll 1, \forall z$.

Note that this assumption is may not satisfied in actual systems. However, as we pointed out at the start of this section, this is not a problem, because the closed-form solution that we are discussing in this section is only used to draw inspiration to write a more general and effective formula, as discussed in the next section.

With the above assumption, since $N > 1$, we have also $NP_0 \times \frac{C_{R,max}}{\Delta f_{ISRS}} \times \Delta f_{ch} \times \frac{1-\exp(-2\alpha_0 z)}{2\alpha_0} \ll 1$, $\forall z$ and therefore we can also write the below approximation:

$$\frac{\sinh\left(\frac{NP_0}{2} \times \frac{C_{R,max}}{\Delta f_{ISRS}} \times \Delta f_{ch} \times \frac{1-\exp(-2\alpha_0 z)}{2\alpha_0}\right)}{\sinh\left(\frac{N^2 P_0}{2} \times \frac{C_{R,max}}{\Delta f_{ISRS}} \times \Delta f_{ch} \times \frac{1-\exp(-2\alpha_0 z)}{2\alpha_0}\right)} \cong \frac{1}{N} \qquad eq.~(7)$$

Therefore, assuming the seven conditions mentioned above, the power evolution for each WDM channel can be approximately written as [21]:

$$P_j(z) \cong P_0 \times e^{-2\alpha_0 z} \times \left\{ e^{\frac{NP_0}{2} \times \frac{C_{R,max}}{\Delta f_{ISRS}} \times (f_N + f_1 - 2f_j) \times \frac{1-\exp(-2\alpha_0 z)}{2\alpha_0}} \right\} \qquad eq.~(8)$$

Alternatively, we can write eq. (8) as:

$$P_j(z) \cong P_j(0) \times e^{-2\alpha_{0,j}z + \frac{2\alpha_{1,j}}{\sigma_j}(\exp(-\sigma_j z)-1)} \qquad eq.~(9)$$

where in eq. (9) we have used:

$$\alpha_{0,j} \triangleq \alpha_0 \qquad eq.~(10)$$

$$\alpha_{1,j} \triangleq \frac{-NP_0}{4} \times \frac{C_{R,max}}{\Delta f_{ISRS}} \times (f_N + f_1 - 2f_j) \qquad eq.~(11)$$

$$\sigma_j \triangleq 2\alpha_0 \qquad eq.~(12)$$

As a conclusion for this section, the power evolution of each channel in the WDM comb can be analytically modeled as eq. (9) provided that the seven conditions mentioned in this section hold.

As we see from equations (10) and (12), $\alpha_{0,j}$ and $\sigma_j$ are the same for all channels and do not depend on $j$ (they are independent of frequency) while from eq. (11) it is obvious that $\alpha_{1,j}$ depends on $j$ (depends on frequency) and therefore it is different channel by channel in the WDM comb.

## 4- Proposed Approximate Model for WDM Channel Power Evolution in the Presence of SRS

In the previous section, we saw a closed-form solution under several restricting conditions which resulted in the power evolution model presented in eq. (9). The aforementioned restricting conditions limit eq. (9) applications and do not provide a general solution for arbitrary scenarios. For example, when we have frequency dependent loss or nonuniform launch power for different channels, eq. (9) is not applicable due to the assumptions made for its derivation. Also, condition 7 is often not met in practical systems.

However, drawing inspiration from the analytical result eq. (9), we consider an *approximate model* for power evolution, which bears a resemblance with eq. (9) but is more flexible. It was proposed in [11]:

$$\begin{aligned} P(f,z) &\cong P(f,0) \times e^{-2\alpha_0(f)z + \frac{2\alpha_1(f)}{\sigma(f)}(\exp(-\sigma(f)z)-1)} \\ &= P(f,0) \times e^{-2\int_0^z (\alpha_0(f) + \alpha_1(f) \times \exp(-\sigma(f)z'))dz'} \end{aligned} \qquad eq.~(13)$$

In eq. (13), $\alpha_0(f)$, $\alpha_1(f)$ and $\sigma(f)$ are three constant numbers (with respect to $z$) which depend on the center frequency of the channel whose power evolution is modeled. *Their value can therefore be different channel by channel.*

The rationale behind the approximate model eq. (13) is twofold. First, though actual systems are in general not compliant with assumptions 1-7 listed before, we assume that the general behavior of the SRS effect will not differ too much from the solution eq (9).

introducing the possibility of channel-dependent $\alpha_0$, $\alpha_1$ and $\sigma$, one can hope to capture the different behavior of a less ideal system configuration. Second, the model eq. (13) is still simple enough so that it can be placed into the GN-model equations to obtain a closed-form solution for system NLI.

To be able to use the power evolution eq. (13) within an NLI model, we need to make sure that we can find the three parameters $\alpha_0(f)$, $\alpha_1(f)$ and $\sigma(f)$, for each channel, in such a way that eq. (13) approximates their actual power evolution accurately enough.

In the following, we discuss how to optimally estimate the above three parameters. First, we assume that we have obtained the actual 'true' power evolution $P_{num}(f, z)$ accurately, for instance by means of numerical integration of eq. (1). In general, we could then always write:

$$P_{num}(f,z) = P(f,0) \times e^{-2\alpha_0(f)z + \frac{2\alpha_1(f)}{\sigma(f)}(\exp(-\sigma(f)z)-1)} + \varepsilon(f,z) \qquad eq.~(14)$$

where $\varepsilon(f,z)$ would be the error that is incurred between the model eq. (13) and the true power evolution. From eq. (14) we have:

$$P_{num}(f,z) - \varepsilon(f,z) = P(f,0) \times e^{-2\alpha_0(f)z + \frac{2\alpha_1(f)}{\sigma(f)}(\exp(-\sigma(f)z)-1)} \qquad eq.~(15)$$

Also, we see from eq. (15):

$$\frac{P_{num}(f,z) - \varepsilon(f,z)}{P(f,0)} = e^{-2\alpha_0(f)z + \frac{2\alpha_1(f)}{\sigma(f)}(\exp(-\sigma(f)z)-1)} \qquad eq.~(16)$$

Taking the natural logarithm of both sides of eq. (16) we get:

$$\ln\{P_{num}(f,z) - \varepsilon(f,z)\} = \ln(P(f,0)) - 2\alpha_0(f)z - \frac{2\alpha_1(f)}{\sigma(f)} + \frac{2\alpha_1(f)}{\sigma(f)}\exp(-\sigma(f)z) \qquad eq.~(17)$$

Also, we can write:

$$\ln\left\{P_{num}(f,z) \times \left(1 - \frac{\varepsilon(f,z)}{P_{num}(f,z)}\right)\right\} = \ln(P(f,0)) - 2\alpha_0(f)z - \frac{2\alpha_1(f)}{\sigma(f)} + \frac{2\alpha_1(f)}{\sigma(f)}\exp(-\sigma(f)z) \qquad eq.~(18)$$

Therefore:

$$ln\{P_{num}(f,z)\} + ln\left\{(1 - \frac{\varepsilon(f,z)}{P_{num}(f,z)})\right\}$$
$$= ln(P(f,0)) - 2\alpha_0(f)z - \frac{2\alpha_1(f)}{\sigma(f)} + \frac{2\alpha_1(f)}{\sigma(f)}\exp(-\sigma(f)z) \qquad eq.\ (19)$$

Now we assume that $|\varepsilon(f,z)| \ll |P_{num}(f,z)|$, therefore $\left|\frac{\varepsilon(f,z)}{P_{num}(f,z)}\right| \ll 1$ and based on the first terms of the Taylor expansion of $ln(.)$ we can write:

$$ln\left\{(1 - \frac{\varepsilon(f,z)}{P_{num}(f,z)})\right\} \cong -\frac{\varepsilon(f,z)}{P_{num}(f,z)} \qquad eq.\ (20)$$

Therefore, from eq. (20) we can write eq. (19) as:

$$ln\{P_{num}(f,z)\} - \frac{\varepsilon(f,z)}{P_{num}(f,z)}$$
$$\cong ln(P(f,0)) - 2\alpha_0(f)z - \frac{2\alpha_1(f)}{\sigma(f)} + \frac{2\alpha_1(f)}{\sigma(f)}\exp(-\sigma(f)z) \qquad eq.\ (21)$$

And simplifying eq. (21) we have:

$$\varepsilon(f,z) \cong P_{num}(f,z)$$
$$\times \left\{ ln\left\{\frac{P_{num}(f,z)}{P(f,0)}\right\} + 2\alpha_0(f)z + \frac{2\alpha_1(f)}{\sigma(f)} - \frac{2\alpha_1(f)}{\sigma(f)}\exp(-\sigma(f)z)\right\} \qquad eq.\ (21)$$

In general, $\varepsilon(f,z) \neq 0$. The goal is that of finding the parameters $\alpha_0(f)$, $\alpha_1(f)$ and $\sigma(f)$ that make such error as small as possible. However, $\varepsilon(f,z)$ is a function of $z$ and therefore we need some cost function that looks at the error over the span length. One possibility would be:

$$f_{Cost} \triangleq \int_0^{L_s} \left(\frac{\varepsilon(f,z)}{P_{num}(f,z)}\right)^2 dz \qquad eq.\ (22)$$

where $L_s$ is the length of the fiber span.

However, we are using the model eq. (13) for nonlinearity evaluation in the fiber. We know that the higher the power, the higher the nonlinearity. So, we would like to attribute a weight to the cost function eq. (22) in such a way that it gives the error a higher weight when the power is higher. One possibility is to modify the cost function presented in eq. (22) as:

$$f_{Cost} \triangleq \int_0^{L_s} \left(P_{num}(f,z)\right)^{m_c} \times \left(\frac{\varepsilon(f,z)}{P_{num}(f,z)}\right)^2 dz \qquad eq.\ (23)$$

where $m_c \geq 0$ is a constant and can be optionally selected to have a better estimation.

Combining equations (21) and (23), the cost function is written as:

$$f_{Cost} \cong \int_0^{L_s} (P_{num}(f,z))^{m_c} \qquad \text{eq. (24)}$$
$$\times \left( \ln\left\{\frac{P_{num}(f,z)}{P(f,0)}\right\} + 2\alpha_0(f)z + 2\alpha_1(f) \times \left\{\frac{1-\exp(-\sigma(f)z)}{\sigma(f)}\right\} \right)^2 dz$$

Eq. (24) is the basis for our estimation of parameters $\alpha_0(f)$, $\alpha_1(f)$ and $\sigma(f)$. For an accurate estimation, the cost function must be minimized. To do this, we assume that the value of the parameter $\sigma(f)$ is known and based on this assumption, we find the optimum values of $\alpha_0(f)$ and $\alpha_1(f)$. To find such optimum values of $\alpha_0(f)$ and $\alpha_1(f)$ we should have:

$$\frac{\partial f_{Cost}}{\partial \alpha_0} = \frac{\partial f_{Cost}}{\partial \alpha_1} = 0 \qquad \text{eq. (25)}$$

Combining equations (24) and (25) we find the two below formulas:

$$\frac{\partial f_{Cost}}{\partial \alpha_0} \cong \int_0^{L_s} (P_{num}(f,z))^{m_c} \times 4 \times z \times \left( \ln\left\{\frac{P_{num}(f,z)}{P(f,0)}\right\} + 2\alpha_0(f)z \right. \qquad \text{eq. (26)}$$
$$\left. + 2\alpha_1(f) \times \left\{\frac{1-\exp(-\sigma(f)z)}{\sigma(f)}\right\} \right) \times dz = 0$$

$$\frac{\partial f_{Cost}}{\partial \alpha_1} \cong \int_0^{L_s} (P_{num}(f,z))^{m_c} \times 4 \times \left\{\frac{1-\exp(-\sigma(f)z)}{\sigma(f)}\right\} \times \left( \ln\left\{\frac{P_{num}(f,z)}{P(f,0)}\right\} \right. \qquad \text{eq. (27)}$$
$$\left. + 2\alpha_0(f)z + 2\alpha_1(f) \times \left\{\frac{1-\exp(-\sigma(f)z)}{\sigma(f)}\right\} \right) \times dz = 0$$

Then eqs. (26) and (27) can be rewritten as:

$$\alpha_0(f) \times \int_0^{L_s} (P_{num}(f,z))^{m_c} \times z^2 dz \qquad \text{eq. (28)}$$
$$+ \alpha_1(f) \times \int_0^{L_s} (P_{num}(f,z))^{2+m_c} \times z \times \left\{\frac{1-\exp(-\sigma(f)z)}{\sigma(f)}\right\} dz$$
$$= -\frac{1}{2} \times \int_0^{L_s} (P_{num}(f,z))^{2+m_c} \times z \times \ln\left\{\frac{P_{num}(f,z)}{P(f,0)}\right\} dz$$

$$\alpha_0(f) \times \int_0^{L_s} (P_{num}(f,z))^{m_c} \times z \times \left\{\frac{1-\exp(-\sigma(f)z)}{\sigma(f)}\right\} dz \qquad \text{eq. (29)}$$

$$+ \alpha_1(f) \times \int_0^{L_s} (P_{num}(f,z))^{2+m_c} \times \left\{\left(\frac{1-\exp(-\sigma(f)z)}{\sigma(f)}\right)^2\right\} dz$$

$$= -\frac{1}{2}$$

$$\times \int_0^{L_s} (P_{num}(f,z))^{2+m_c} \times \left\{\frac{1-\exp(-\sigma(f)z)}{\sigma(f)}\right\} \times \ln\left\{\frac{P_{num}(f,z)}{P(f,0)}\right\} dz$$

Therefore, combining equations (28) and (29), we have the matrix equation:

$$\begin{bmatrix} \int_0^{L_s} (P_{num}(f,z))^{m_c} \times z^2 dz & \int_0^{L_s} (P_{num}(f,z))^{m_c} \times z \times \left\{\frac{1-\exp(-\sigma(f)z)}{\sigma(f)}\right\} dz \\ \int_0^{L_s} (P_{num}(f,z))^{m_c} \times z \times \left\{\frac{1-\exp(-\sigma(f)z)}{\sigma(f)}\right\} dz & \int_0^{L_s} (P_{num}(f,z))^{m_c} \times \left\{\left(\frac{1-\exp(-\sigma(f)z)}{\sigma(f)}\right)^2\right\} dz \end{bmatrix}$$

$$\times \begin{bmatrix} \alpha_0(f) \\ \alpha_1(f) \end{bmatrix} = \begin{bmatrix} -\frac{1}{2} \times \int_0^{L_s} (P_{num}(f,z))^{m_c} \times z \times \ln\left\{\frac{P_{num}(f,z)}{P(f,0)}\right\} dz \\ -\frac{1}{2} \times \int_0^{L_s} (P_{num}(f,z))^{m_c} \times \left\{\frac{1-\exp(-\sigma(f)z)}{\sigma(f)}\right\} \times \ln\left\{\frac{P_{num}(f,z)}{P(f,0)}\right\} dz \end{bmatrix}$$

eq. (30.1)

As a result, we get:

$$\begin{bmatrix} \alpha_0(f) \\ \alpha_1(f) \end{bmatrix} = \qquad \text{eq. (30.2)}$$

$$\begin{bmatrix} \int_0^{L_s} (P_{num}(f,z))^{m_c} \times z^2 dz & \int_0^{L_s} (P_{num}(f,z))^{m_c} \times z \times \left\{\frac{1-\exp(-\sigma(f)z)}{\sigma(f)}\right\} dz \\ \int_0^{L_s} (P_{num}(f,z))^{m_c} \times z \times \left\{\frac{1-\exp(-\sigma(f)z)}{\sigma(f)}\right\} dz & \int_0^{L_s} (P_{num}(f,z))^{m_c} \times \left\{\left(\frac{1-\exp(-\sigma(f)z)}{\sigma(f)}\right)^2\right\} dz \end{bmatrix}^{-1}$$

$$\times \begin{bmatrix} -\frac{1}{2} \times \int_0^{L_s} (P_{num}(f,z))^{m_c} \times z \times ln\left\{\frac{P_{num}(f,z)}{P(f,0)}\right\} dz \\ -\frac{1}{2} \times \int_0^{L_s} (P_{num}(f,z))^{m_c} \times \left\{\frac{1-\exp(-\sigma(f)z)}{\sigma(f)}\right\} \times ln\left\{\frac{P_{num}(f,z)}{P(f,0)}\right\} dz \end{bmatrix}$$

So, provided that $\sigma(f)$ is known, the optimum values for $\alpha_0(f)$ and $\alpha_1(f)$ are directly calculated in closed form based on eq. (30).

For finding $\sigma(f)$ we can use a search algorithm like the Golden section search method with a starting point of $\sigma(f) = 2 \times \alpha(f)$ where $\alpha(f)$ is the fiber loss parameter in the absence of SRS at the frequency $f$ (if $f = f_j$; $\alpha(f) = \alpha_j$), drawing inspiration from eq. (12). In each iteration of the search, a fixed value for $\sigma(f)$ is assumed; afterwards $\alpha_0(f)$ and $\alpha_1(f)$ are calculated based on eq. (30). Having $\sigma(f)$, $\alpha_0(f)$ and $\alpha_1(f)$, the cost function is calculated through eq. (24) and is compared with cost function values in previous iterations. By following this procedure and searching in a reasonable interval, i.e., $\sigma(f) \epsilon [\alpha(f) \ \ 4 \times \alpha(f)]$, we can approach the optimum value of $\sigma(f)$ by searching iteratively.

## 5- Nonlinear interference modeling using the incoherent GN model

In the following, we will report the detailed derivation of a new CFM whose features extend beyond those of [11]. In the first part of this section, however, we will re-derive the results of [11]. The reason why we report here again the full derivation is that notation is somewhat different and that we deem it convenient for the reader to have the whole derivation procedure available in one place. On the other hand, we also suggest as a preliminary reading [11]. Later in this section, we will depart from [11] to obtain a more advanced result.

The power spectral density (PSD) of the nonlinear interference (NLI) based on the general formula of the GN model is [3]:

$$G_{NLI}(f) = \frac{16}{27} \int_{-\infty}^{+\infty} \int_{-\infty}^{+\infty} G_s(f_1) G_s(f_2) G_s(f_1 + f_2 - f) \times |LK(f_1, f_2, f_1 + f_2 - f)|^2 \, df_1 df_2 \qquad eq.\ (31)$$

where in eq. (31), $G_s(f)$ is the power spectral density (PSD) of the WDM signal launched into the fiber and $LK$ is the link function which is determined based on the fiber link configuration and will be discussed later. The WDM PSD can be written in terms of each channel PSD as:

$$G_s(f) = \sum_{m_{ch}=1}^{N_c} G_{m_{ch}}(f) \qquad eq.\ (32)$$

where $N_c$ is the number of channels available in WDM comb and $G_{m_{ch}}(f)$ is the PSD due to $m_{ch}$'th channel in the WDM comb.

Combining eq. (31) and eq. (32) we will have:

$$G_{NLI}(f) = \frac{16}{27} \sum_{m_{ch}=1}^{N_c} \sum_{n_{ch}=1}^{N_c} \sum_{k_{ch}=1}^{N_c} \int_{-\infty}^{+\infty} \int_{-\infty}^{+\infty} G_{m_{ch}}(f_1) G_{n_{ch}}(f_2)$$
$$\times G_{k_{ch}}(f_1 + f_2 - f) |LK(f_1, f_2, f_1 + f_2 - f)|^2 \, df_1 df_2$$

eq. (33)

In general, the power spectral density (PSD) of each channel in the WDM comb is raised-cosine with nonzero roll-off. For now, we approximately replace it with a rectangular channel with the same center frequency as the original raised cosine channel. Also, we assume that the null-to-null bandwidth of the approximate rectangular channel is equal to the symbol rate (baud rate) of the original raised cosine channel. We set the flat-top value of the PSD of the rectangle function to the same value as that of the flat-top of the PSD of the original raised cosine channel, as it is shown in Figure (1).

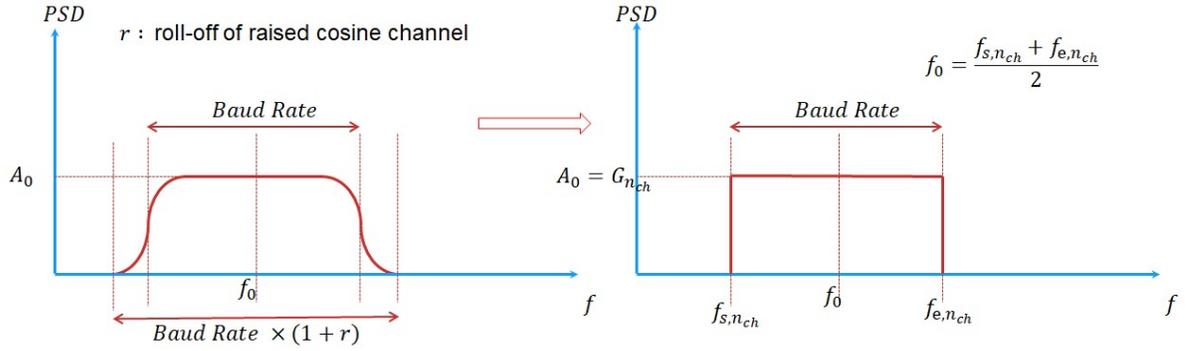

Figure (1): Replacing a raised cosine channel with a rectangular channel

Therefore, we have:

$$G_{m_{ch}}(f) = \begin{cases} G_{m_{ch}} & f_{s,m_{ch}} \leq f \leq f_{e,m_{ch}} \\ 0 & otherwise \end{cases}$$

eq. (34)

where in eq. (34), $f_{s,m_{ch}}$ and $f_{e,m_{ch}}$ are the start and end frequency of the $m_{ch}$'th channel in WDM comb respectively. Also $G_{m_{ch}}$ is the constant value of PSD due to the $m_{ch}$'th channel.

By using eq. (34), eq. (33) can be written as:

$$G_{NLI}(f) = \frac{16}{27} \sum_{m_{ch}=1}^{N_c} \sum_{n_{ch}=1}^{N_c} \sum_{k_{ch}=1}^{N_c} G_{m_{ch}} G_{n_{ch}} \int_{f_{s,n_{ch}}}^{f_{e,n_{ch}}} \int_{f_{s,m_{ch}}}^{f_{e,m_{ch}}} G_{k_{ch}}(f_1 + f_2 - f)$$
$$|LK(f_1, f_2, f_1 + f_2 - f)|^2 \, df_1 df_2$$

eq. (35)

As assumed in eq. (34), $G_n(f) \, \forall n$ has rectangular shape. As a result, eq. (35) can be written as:

$$G_{NLI}(f) = \frac{16}{27} \sum_{m_{ch}=1}^{N_c} \sum_{n_{ch}=1}^{N_c} \sum_{k_{ch}=1}^{N_c} G_{m_{ch}} G_{n_{ch}} G_{k_{ch}} \qquad \text{eq. (36)}$$

$$\times \iint_{S(m_{ch},n_{ch},k_{ch})} |LK(f_1, f_2, f_1 + f_2 - f)|^2 \, df_1 df_2$$

In eq. (36), $S(m_{ch}, n_{ch}, k_{ch})$ is the 2-D region in $f_1 - f_2$ plane confined by three criteria below:

$$\begin{aligned} f_{s,m_{ch}} &\leq f_1 \leq f_{e,m_{ch}} \\ f_{s,n_{ch}} &\leq f_2 \leq f_{e,n_{ch}} \\ f_{s,k_{ch}} + f &\triangleq f'_{s,k_{ch}} \leq f_1 + f_2 \leq f'_{e,k_{ch}} \triangleq f_{e,k_{ch}} + f \end{aligned} \qquad \text{eq. (37)}$$

We call the region defined by eq. (37) criteria, an *integration island*. In figure (2), we can see a typical plot of the $S(m_{ch}, n_{ch}, k_{ch})$, hatched by blue color, in the $f_1 - f_2$ plane.

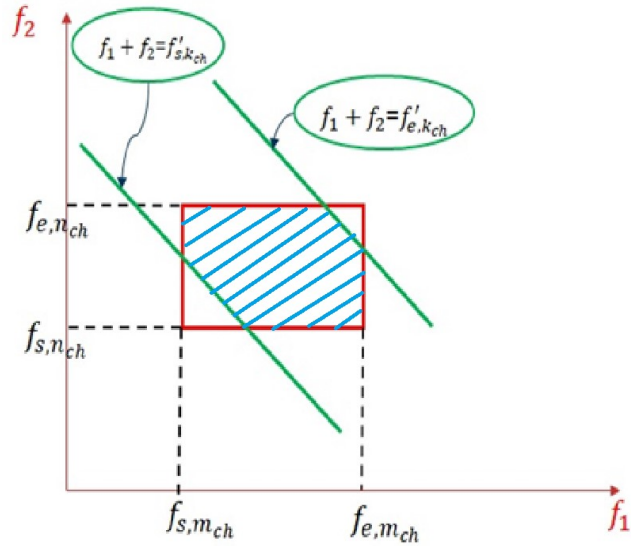

*Figure (2): The scheme of the formation of a typical Integration island in the f1-f2 plane*

It is worth noting that in eq. (36), each of the parameters $m_{ch}, n_{ch}$ and $k_{ch}$ can vary from 1 to $N_c$ and in general we have $N_c^3$ integration islands. However, many of these islands are null space due to having no overlap of the three criteria in eq. (37), so that their integral is zero.

$$S = \{(m_{ch}, n_{ch}, k_{ch}) \mid \, 1 \leq m_{ch} \leq N_c \,, 1 \leq n_{ch} \leq N_c \,, 1 \leq k_{ch} \leq N_c\} \qquad \text{eq. (38)}$$

Among these $N_c^3$ integration islands, we have the important category of islands called SCI-XCI which is defined as:

$$S^{SCI-XCI} \triangleq \{(m_{ch}, n_{ch}, k_{ch}) \in S \,|\, k_{ch} = n_{ch} \neq CUT, m_{ch} = CUT\} \quad \text{eq. (39)}$$
$$\cup \{(m_{ch}, n_{ch}, k_{ch}) \in S \,|\, k_{ch} = m_{ch} \neq CUT, n_{ch} = CUT\}$$
$$\cup \{(m_{ch}, n_{ch}, k_{ch}) \in S \,|\, k_{ch} = n_{ch} = m_{ch} = CUT\}$$

where CUT is the channel under test, i.e., the channel located within the interval: ($f_{s,CUT} \leq f \leq f_{e,CUT}$). When we are not working in zero or very low fiber dispersion regime, we can approximately replace $S$ by $S^{SCI-XCI}$ in eq. (36) [11]. This is because high dispersion makes the contribution of multichannel interference (MCI) islands negligible [11]. If so, we will have:

$$G_{NLI}(f) \cong \frac{16}{27} \times \sum_{\substack{n_{ch}=1 \\ n_{ch} \neq CUT}}^{N_c} G_{CUT} G_{n_{ch}}^2 \times \iint\limits_{S(CUT, n_{ch}, n_{ch})} |LK(f_1, f_2, f_1 + f_2 - f)|^2 \, df_1 df_2 \quad \text{eq. (40)}$$

$$+ \frac{16}{27} \times \sum_{\substack{m_{ch}=1 \\ m_{ch} \neq CUT}}^{N_c} G_{CUT} G_{m_{ch}}^2 \times \iint\limits_{S(m_{ch}, CUT, m_{ch})} |LK(f_1, f_2, f_1 + f_2 - f)|^2 \, df_1 df_2$$

$$+ \frac{16}{27} \times G_{CUT}^3 \times \iint\limits_{S(CUT, CUT, CUT)} |LK(f_1, f_2, f_1 + f_2 - f)|^2 \, df_1 df_2$$

Due to symmetric behavior $LK(f_1, f_2, f_1 + f_2 - f)$ function with respect to it's two variables $f_1, f_2$, eq. (40) can be rewritten compactly as:

$$G_{NLI}(f) \cong \frac{16}{27} \times \sum_{m_{ch}=1}^{N_c} G_{CUT} G_{m_{ch}}^2 \times (2 - \delta_{CUT, m_{ch}}) \quad \text{eq. (41)}$$
$$\times \iint\limits_{S(CUT, m_{ch}, m_{ch})} |LK(f_1, f_2, f_1 + f_2 - f)|^2 \, df_1 df_2$$

where $\delta_{i,j}$ is the *Kronecker delta function* ($\delta_{i,j} = 1 \text{ for } i = j$ and $\delta_{i,j} = 0 \text{ for } i \neq j$). Also, we concentrate on the calculation of the NLI at the center frequency of the channel under test that we call $f_{CUT}$. We set $f = f_{c,CUT} \triangleq f_{CUT}$ and obtain:

$$G_{NLI}(f_{CUT}) \cong \frac{16}{27} \qquad \text{eq. (42)}$$
$$\times \sum_{m_{ch}=1}^{N_c} G_{CUT} G_{m_{ch}}^2 \times (2 - \delta_{CUT,m_{ch}})$$
$$\times \iint_{S(CUT,m_{ch},m_{ch})} |LK(f_1, f_2, f_1 + f_2 - f_{CUT})|^2 \, df_1 df_2$$

The double integral in eq. (42) can be approximated as below by replacing a complex shape integration island with a rectangular shape island [11]:

$$\iint_{S(CUT,m_{ch},m_{ch})} |LK(f_1, f_2, f_1 + f_2 - f_{CUT})|^2 \, df_1 df_2 \qquad \text{eq. (43)}$$
$$\cong \int_{f_{s,CUT}}^{f_{e,CUT}} \int_{f_{s,m_{ch}}}^{f_{e,m_{ch}}} |LK(f_1, f_2, f_1 + f_2 - f_{CUT})|^2 \, df_1 df_2$$

Using eq. (43), eq. (42) can be rewritten as:

$$G_{NLI}(f_{CUT}) \cong \frac{16}{27} \qquad \text{eq. (44)}$$
$$\times \sum_{m_{ch}=1}^{N_c} G_{CUT} G_{m_{ch}}^2 \times (2 - \delta_{CUT,m_{ch}})$$
$$\times \int_{f_{s,CUT}}^{f_{e,CUT}} \int_{f_{s,m_{ch}}}^{f_{e,m_{ch}}} |LK(f_1, f_2, f_1 + f_2 - f_{CUT})|^2 \, df_1 df_2$$

Now we investigate the link function available in the GN model formula. The link function in eq. (44) can be expressed as [18],[22]:

$$LK(f_1, f_2, f_3) =$$
$$-j \sum_{n_s=1}^{N_s} \gamma_{n_s} \times \left\{ \int_0^{L_s(n_s)} e^{\int_0^{z'} [\kappa_{n_s}(z'', f_1) + \kappa_{n_s}^*(z'', f_3) + \kappa_{n_s}(z'', f_2) - \kappa_{n_s}(z'', f_1 + f_2 - f_3)] \, dz''} \, dz' \right\} \qquad \text{eq. (45)}$$

$$\times \left\{ \prod_{p=n_s}^{N_s} \Gamma_p^{\frac{1}{2}}(f_1+f_2-f_3)\, e^{j\theta_p(f_1+f_2-f_3)}\, e^{\int_0^{L_s(p)} \kappa_p(z,f_1+f_2-f_3)\,dz}\, e^{-j\beta_{DCU}^{(p)}(f_1+f_2-f_3)} \right\}$$

$$\times \left\{ \prod_{p=1}^{n_s-1} \left\{ [\Gamma_p(f_1)\Gamma_p(f_2)\Gamma_p(f_3)]^{\frac{1}{2}} \times e^{\int_0^{L_s(p)}[\kappa_p(z,f_1)+\kappa_p(z,f_2)+\kappa_p^*(z,f_3)]\,dz} \times \right. \right.$$

$$\left. \left. e^{j[\theta_p(f_1)+\theta_p(f_2)-\theta_p(f_3)]} \times e^{-j\left[\beta_{DCU}^{(p)}(f_1)+\beta_{DCU}^{(p)}(f_2)-\beta_{DCU}^{(p)}(f_3)\right]} \right\} \right\}$$

where $L_s(n_s)$ is the physical length of the $n_s'th$ fiber span, $N_s$ is the number of fiber spans in the link, $\gamma_{n_s}$ is the nonlinearity parameter of the $n_s'th$ fiber span which is assumed to be a constant with respect to z (distance) and f (frequency) in one span but it can be different constant values span by span. The EDFA at the end of each span is modeled by an ideal flat gain amplifier cascaded with a linear time invariant (LTI) filter. $\Gamma_{n_s}(f)$ is the frequency dependent power gain of the actual EDFA (ideal flat-gain EDFA+LTI filter) at the end of the $n_s'th$ fiber span and $\theta_{n_s}(f)$ is the frequency dependent phase imposed to the electric field through the real EDFA (ideal EDFA+LTI filter). Therefore, the input-output relation of the electrical field for the real EDFA (ideal EDFA+LTI filter) is modeled as $E_{out}(f) = \sqrt{\Gamma_{n_s}(f)} \times e^{+j\theta_{n_s}(f)} \times E_{in}(f)$ where $E_{in}(f)$ is the input electrical field (in frequency domain) at the end of fiber span $n_s$ which enters the real EDFA (ideal EDFA+LTI filter) and $E_{out}(f)$ will be the electrical field (in frequency domain) at the output of real EDFA (ideal EDFA+LTI filter) which is in fact the electric field at the start of span $n_s+1$. $\beta_{DCU}^{(n_s)}(f)$ is lumped accumulated dispersion at the end of the $n_s'th$ fiber span. $\kappa_{n_s}(z,f)$ is a complex valued function, called the generalized propagation constant for the $n_s'th$ fiber span, which is defined as [6],[22]:

$$\kappa_{n_s}(z,f) \triangleq -j\beta_{n_s}(z,f) - \alpha_{n_s}(z,f) \qquad eq.\ (46)$$

In fact, $Real\{\kappa_{n_s}(z,f)\} = -\alpha_{n_s}(z,f)$ and $Imag\{\kappa_{n_s}(z,f)\} = -\beta_{n_s}(z,f)$ where $\alpha_{n_s}(z,f)$ and $\beta_{n_s}(z,f)$ are two real valued functions. $\beta_{n_s}(z,f)$ is called propagation constant and is assumed to be independent of z (constant along one fiber span) and can be expressed as:

$$\beta_{n_s}(z,f) = \beta_{n_s}(f) = \beta_{0,n_s} + 2\pi\beta_{1,n_s}(f - f_{n_s}^c) + 4\pi^2 \frac{\beta_{2,n_s}}{2}(f - f_{n_s}^c)^2 + \qquad eq.\ (47)$$
$$8\pi^3 \frac{\beta_{3,n_s}}{6}(f - f_{n_s}^c)^3$$

In equation (47), $f_{n_s}^c$ is the center frequency for the Taylor expansion and in general can be different span by span but over each span it must be a constant. Also $\beta_{0,n_s}$, $\beta_{1,n_s}$, $\beta_{2,n_s}$, $\beta_{3,n_s}$ are constant values along each span but they can change span by span.

Also $\alpha_{n_s}(z, f)$ in eq. (46) is called loss function and it is assumed to be a slowly varying function with respect to frequency. We assume the loss function to be approximately constant over each WDM channel and equal to its value at the center frequency of the WDM channel:

$$\alpha_{n_s}(z, f) \cong \alpha_{n_s}(z, f_{c,m_{ch}}) \qquad f_{s,m_{ch}} \leq f \leq f_{e,m_{ch}} \qquad \text{eq. (48)}$$

where $f_{c,m_{ch}}$ is the center frequency of the $m_{ch}$ channel which is $f_{c,m_{ch}} \triangleq \frac{f_{s,m_{ch}} + f_{e,m_{ch}}}{2}$. If we are looking at the channel under test, which as mentioned we call CUT, we write $f_{c,CUT} \triangleq f_{CUT} = \frac{f_{s,CUT} + f_{e,CUT}}{2}$).

Therefore, when $f_{s,m_{ch}} \leq f_1 \leq f_{e,m_{ch}}$ and $f_{s,CUT} \leq f_2 \leq f_{e,CUT}$, having equations (46) and (48) we can write:

$$\kappa_{n_s}(z'', f_1) + \kappa_{n_s}^*(z'', f_1 + f_2 - f_{CUT}) + \kappa_{n_s}(z'', f_2) - \kappa_{n_s}(z'', f_{CUT}) = \qquad \text{eq. (49)}$$
$$-\alpha_{n_s}(z'', f_1) - \alpha_{n_s}(z'', f_1 + f_2 - f_{CUT}) - \alpha_{n_s}(z'', f_2) + \alpha_{n_s}(z'', f_{CUT})$$
$$- j\beta_{n_s}(z'', f_1) + j\beta_{n_s}(z'', f_1 + f_2 - f_{CUT}) - j\beta_{n_s}(z'', f_2)$$
$$+ j\beta_{n_s}(z'', f_{CUT}) \cong$$
$$-\alpha_{n_s}(z'', f_{c,m_{ch}}) - \alpha_{n_s}(z'', f_{c,m_{ch}} + f_{CUT} - f_{CUT}) - \alpha_{n_s}(z'', f_{CUT})$$
$$+ \alpha_{n_s}(z'', f_{CUT}) - j\beta_{n_s}(z'', f_1) + j\beta_{n_s}(z'', f_1 + f_2 - f_{CUT})$$
$$- j\beta_{n_s}(z'', f_2) + j\beta_{n_s}(z'', f_{CUT}) =$$
$$-2\alpha_{n_s}(z'', f_{c,m_{ch}}) - j\beta_{n_s}(z'', f_1) + j\beta_{n_s}(z'', f_1 + f_2 - f_{CUT}) - j\beta_{n_s}(z'', f_2) +$$
$$j\beta_{n_s}(z'', f_{CUT})$$

$$\text{; for } f_{s,m_{ch}} \leq f_1 \leq f_{e,m_{ch}} \text{ and } f_{s,CUT} \leq f_2 \leq f_{e,CUT}$$

Also, we can make eq. (49) simpler by using eq. (47) as:

$$-j\beta_{n_s}(z'', f_1) + j\beta_{n_s}(z'', f_1 + f_2 - f_{CUT}) - j\beta_{n_s}(z'', f_2) + j\beta_{n_s}(z'', f_{CUT}) = \qquad \text{eq. (50)}$$
$$j4\pi^2 (f_1 - f)(f_2 - f) \times \left(\beta_{2,n_s} + \pi\beta_{3,n_s}(f_1 + f_2 - 2f_{n_s}^c)\right)$$

Therefore using (50), eq. (49) is simplified as:

$$\kappa_{n_s}(z'', f_1) + \kappa_{n_s}^*(z'', f_1 + f_2 - f_{CUT}) + \kappa_{n_s}(z'', f_2) - \kappa_{n_s}(z'', f_{CUT}) \cong \qquad \text{eq. (50.2)}$$
$$-2\alpha_{n_s}(z'', f_{c,m_{ch}}) + j4\pi^2 (f_1 - f)(f_2 - f) \times \left(\beta_{2,n_s} + \pi\beta_{3,n_s}(f_1 + f_2 - 2f_{n_s}^c)\right)$$
$$\text{;for } f_{s,m_{ch}} \leq f_1 \leq f_{e,m_{ch}} \text{ and } f_{s,CUT} \leq f_2 \leq f_{e,CUT}$$

We can then write:

$$e^{\int_0^{z'} [\kappa_{n_s}(z'', f_1) + \kappa_{n_s}^*(z'', f_3) + \kappa_{n_s}(z'', f_2) - \kappa_{n_s}(z'', f_1 + f_2 - f_3)] dz''} \cong \qquad \text{eq. (51)}$$

$$e^{\int_0^{z'}\left[-2\alpha_{n_s}(z'',f_{c,m_{ch}})+j4\pi^2(f_1-f)(f_2-f)\times\left(\beta_{2,n_s}+\pi\beta_{3,n_s}(f_1+f_2-2f_{n_s}^C)\right)\right]dz''}$$

$$= e^{-2\int_0^{z'}\alpha_{n_s}(z'',f_{c,m_{ch}})dz''} \times e^{jz'\times 4\pi^2(f_1-f)(f_2-f)\times\left(\beta_{2,n_s}+\pi\beta_{3,n_s}(f_1+f_2-2f_{n_s}^C)\right)}$$

$$; \text{ for } f_{s,m_{ch}} \le f_1 \le f_{e,m_{ch}} \text{ and } f_{s,CUT} \le f_2 \le f_{e,CUT}$$

Therefore using eq. (51) we have:

eq. (52)
$$\int_0^{L_s(n_s)} e^{\int_0^{z'}\left[\kappa_{n_s}(z'',\,f_1)+\kappa_{n_s}^*(z'',f_3)+\kappa_{n_s}(z'',f_2)-\kappa_{n_s}(z'',f_1+f_2-f_3)\right]dz''} dz'$$

$$= \int_0^{L_s(n_s)} e^{-2\int_0^{z'}\alpha_{n_s}(z'',f_{c,m_{ch}})dz''} \times e^{jz'\times 4\pi^2(f_1-f)(f_2-f)\times\left(\beta_{2,n_s}+\pi\beta_{3,n_s}(f_1+f_2-2f_{n_s}^C)\right)} dz'$$

$$; \text{ for } f_{s,m_{ch}} \le f_1 \le f_{e,m_{ch}} \text{ and } f_{s,CUT} \le f_2 \le f_{e,CUT}$$

We now use the model presented in eq. (13) for the loss function. We assume:

$$\alpha_{n_s}(z,f) = \alpha_0^{(n_s)}(f) + \alpha_1^{(n_s)}(f)\exp\left(-\sigma^{(n_s)}(f)z\right)$$

eq. (53)

With eq. (53), we can rewrite eq. (52) as:

eq. (54)
$$\int_0^{L_s(n_s)} e^{\int_0^{z'}\left[\kappa_{n_s}(z'',\,f_1)+\kappa_{n_s}^*(z'',f_3)+\kappa_{n_s}(z'',f_2)-\kappa_{n_s}(z'',f_1+f_2-f_3)\right]dz''} dz'$$

$$= \int_0^{L_s(n_s)} e^{\left\{-2\alpha_0^{(n_s)}(f_{c,m_{ch}})z' - \frac{2\alpha_1^{(n_s)}(f_{c,m_{ch}})}{\sigma^{(n_s)}(f_{c,m_{ch}})} + \frac{2\alpha_1^{(n_s)}(f_{c,m_{ch}})}{\sigma^{(n_s)}(f_{c,m_{ch}})}\times\exp\left(-\sigma^{(n_s)}(f_{c,m_{ch}})z'\right)\right\}}$$

$$\times e^{jz'\times 4\pi^2(f_1-f)(f_2-f)\times\left(\beta_{2,n_s}+\pi\beta_{3,n_s}(f_1+f_2-2f_{n_s}^C)\right)} dz' =$$

$$e^{-\frac{2\alpha_1^{(n_s)}(f_{c,m_{ch}})}{\sigma^{(n_s)}(f_{c,m_{ch}})}}$$

$$\times \int_0^{L_s(n_s)} e^{\left\{\frac{2\alpha_1^{(n_s)}(f_{c,m_{ch}})}{\sigma^{(n_s)}(f_{c,m_{ch}})}\times\exp\left(-\sigma^{(n_s)}(f_{c,m_{ch}})z'\right)\right\}}$$

$$\times e^{z'\times\left\{-2\alpha_0^{(n_s)}(f_{c,m_{ch}})+j\times 4\pi^2(f_1-f)(f_2-f)\times\left(\beta_{2,n_s}+\pi\beta_{3,n_s}(f_1+f_2-2f_{n_s}^C)\right)\right\}} dz'$$

$$; \text{ for } f_{s,m_{ch}} \leq f_1 \leq f_{e,m_{ch}} \text{ and } f_{s,CUT} \leq f_2 \leq f_{e,CUT}$$

So far, we have re-derived the results based on [11]. From now on we use a different procedure. Based on a Taylor expansion series we can write:

$$e^{\left\{\frac{2\alpha_1^{(n_s)}(f_{c,m_{ch}})}{\sigma^{(n_s)}(f_{c,m_{ch}})} \times \exp\left(-\sigma^{(n_s)}(f_{c,m_{ch}})z'\right)\right\}} \quad \text{eq. (55)}$$

$$= \sum_{k=0}^{\infty} \frac{1}{k!} \times \left[\frac{2\alpha_1^{(n_s)}(f_{c,m_{ch}})}{\sigma^{(n_s)}(f_{c,m_{ch}})} \times \exp\left(-\sigma^{(n_s)}(f_{c,m_{ch}})z'\right)\right]^k$$

$$= \sum_{k=0}^{\infty} \frac{1}{k!} \times \left[\frac{2\alpha_1^{(n_s)}(f_{c,m_{ch}})}{\sigma^{(n_s)}(f_{c,m_{ch}})}\right]^k \times \exp\left(-k \times \sigma^{(n_s)}(f_{c,m_{ch}})z'\right)$$

We then assume that we can approximate the infinite summation in eq. (55) with a summation with a finite number of terms, say, $M$. For the finite summation to be accurate enough we should have:

$$\left|\frac{1}{M!} \times \left[\frac{2\alpha_1^{(n_s)}(f_{c,m_{ch}})}{\sigma^{(n_s)}(f_{c,m_{ch}})}\right]^M \times \exp\left(-M \times \sigma^{(n_s)}(f_{c,m_{ch}})z'\right)\right| \quad \text{eq. (56)}$$

$$\gg \left|\frac{1}{(M+1)!} \times \left[\frac{2\alpha_1^{(n_s)}(f_{c,m_{ch}})}{\sigma^{(n_s)}(f_{c,m_{ch}})}\right]^{(M+1)} \times \exp\left(-(M+1) \times \sigma^{(n_s)}(f_{c,m_{ch}})z'\right)\right|$$

$$\forall z'$$

We must emphasize that $\sigma^{(n_s)}(f_{c,m_{ch}})$ is always considered as a non-negative real number and the term $\exp(-\sigma^{(n_s)}(f_{c,m_{ch}})z')$ is never bigger than 1. ($0 \leq \exp(-\sigma^{(n_s)}(f_{c,m_{ch}})z') \leq 1$, $\forall z'$). Therefore, manipulating eq. (56) we can get a more conservative condition as:

$$M \gg \left|\frac{2\alpha_1^{(n_s)}(f_{c,m_{ch}})}{\sigma^{(n_s)}(f_{c,m_{ch}})}\right| \quad \text{eq. (57)}$$

Therefore, we could simply choose:

$$M(n_s, f) = 1 + \left\lfloor 10 \times \left|\frac{2\alpha_1^{(n_s)}(f)}{\sigma^{(n_s)}(f)}\right| \right\rfloor \quad \text{eq. (58)}$$

where the sign $\lfloor . \rfloor$ denotes floor function which returns the biggest integer smaller than or equal to its argument.

And eq. (55) can be written as:

$$e^{\left\{\frac{2\alpha_1^{(n_s)}(f_{c,m_{ch}})}{\sigma^{(n_s)}(f_{c,m_{ch}})} \times \exp\left(-\sigma^{(n_s)}(f_{c,m_{ch}})z'\right)\right\}} \qquad eq.\ (59)$$

$$\cong \sum_{k=0}^{M(n_s,f_{c,m_{ch}})} \frac{1}{k!} \times \left[\frac{2\alpha_1^{(n_s)}(f_{c,m_{ch}})}{\sigma^{(n_s)}(f_{c,m_{ch}})}\right]^k \times \exp\left(-k \times \sigma^{(n_s)}(f_{c,m_{ch}})z'\right)$$

Using eq. (59), eq. (54) is written as:

$$\int_0^{L_s(n_s)} e^{\int_0^{z'}[\kappa_{n_s}(z'',\ f_1)+\kappa_{n_s}^*(z'',f_3)+\kappa_{n_s}(z'',f_2)-\kappa_{n_s}(z'',f_1+f_2-f_3)]\,dz''}\,dz' \cong \qquad eq.\ (60)$$

$$e^{-\frac{2\alpha_1^{(n_s)}(f_{c,m_{ch}})}{\sigma^{(n_s)}(f_{c,m_{ch}})}} \times \sum_{k=0}^{M(n_s,f_{c,m_{ch}})} \frac{1}{k!} \times \left[\frac{2\alpha_1^{(n_s)}(f_{c,m_{ch}})}{\sigma^{(n_s)}(f_{c,m_{ch}})}\right]^k \times$$

$$\int_0^{L_s(n_s)} e^{z' \times \left\{-2\alpha_0^{(n_s)}(f_{c,m_{ch}})-k\times\sigma^{(n_s)}(f_{c,m_{ch}})+j\times 4\pi^2(f_1-f)(f_2-f)\times\left(\beta_{2,n_s}+\pi\beta_{3,n_s}(f_1+f_2-2f_{n_s}^c)\right)\right\}}\,dz'$$

$$;\ for\ f_{s,m_{ch}} \le f_1 \le f_{e,m_{ch}}\ and\ f_{s,CUT} \le f_2 \le f_{e,CUT}$$

Also, the integration in eq. (60) can be solved analytically as:

$$\int_0^{L_s(n_s)} e^{z' \times \left\{-2\alpha_0^{(n_s)}(f_{c,m_{ch}})-k\times\sigma^{(n_s)}(f_{c,m_{ch}})+j\times 4\pi^2(f_1-f)(f_2-f)\times\left(\beta_{2,n_s}+\pi\beta_{3,n_s}(f_1+f_2-2f_{n_s}^c)\right)\right\}}\,dz' \qquad eq.\ (61)$$

$$= \frac{e^{L_s(n_s) \times \left\{-2\alpha_0^{(n_s)}(f_{c,m_{ch}})-k\times\sigma^{(n_s)}(f_{c,m_{ch}})+j\times 4\pi^2(f_1-f)(f_2-f)\times\left(\beta_{2,n_s}+\pi\beta_{3,n_s}(f_1+f_2-2f_{n_s}^c)\right)\right\}} - 1}{-2\alpha_0^{(n_s)}(f_{c,m_{ch}}) - k \times \sigma^{(n_s)}(f_{c,m_{ch}}) + j \times 4\pi^2(f_1-f)(f_2-f) \times \left(\beta_{2,n_s} + \pi\beta_{3,n_s}(f_1+f_2-2f_{n_s}^c)\right)}$$

We assume that $e^{L_s(n_s) \times \left\{-2\alpha_0^{(n_s)}(f_{c,m_{ch}})-k\times\sigma^{(n_s)}(f_{c,m_{ch}})\right\}} \ll 1$ and we will have:

$$\int_0^{L_s(n_s)} e^{z' \times \left\{-2\alpha_0^{(n_s)}(f_{c,m_{ch}})-k\times\sigma^{(n_s)}(f_{c,m_{ch}})+j\times 4\pi^2(f_1-f)(f_2-f)\times\left(\beta_{2,n_s}+\pi\beta_{3,n_s}(f_1+f_2-2f_{n_s}^c)\right)\right\}}\,dz' \qquad eq.\ (62)$$

$$\cong \frac{1}{\left[2\alpha_0^{(n_s)}(f_{c,m_{ch}}) + k \times \sigma^{(n_s)}(f_{c,m_{ch}})\right] - j \times \left[4\pi^2(f_1-f)(f_2-f) \times \left(\beta_{2,n_s} + \pi\beta_{3,n_s}(f_1+f_2-2f_{n_s}^c)\right)\right]}$$

Then using (62), eq. (60) simplifies to:

$$\int_0^{L_S(n_s)} e^{\int_0^{z'}[\kappa_{n_s}(z'', f_1)+\kappa_{n_s}^*(z'',f_3)+\kappa_{n_s}(z'',f_2)-\kappa_{n_s}(z'',f_1+f_2-f_3)] dz''} dz' \cong \quad eq.\ (63)$$

$$e^{-\frac{2\alpha_1^{(n_s)}(f_{c,m_{ch}})}{\sigma^{(n_s)}(f_{c,m_{ch}})}} \times \sum_{k=0}^{M(n_s,f_{c,m_{ch}})} \frac{1}{k!} \times \left[\frac{2\alpha_1^{(n_s)}(f_{c,m_{ch}})}{\sigma^{(n_s)}(f_{c,m_{ch}})}\right]^k \times$$

$$\frac{1}{\left[2\alpha_0^{(n_s)}(f_{c,m_{ch}}) + k \times \sigma^{(n_s)}(f_{c,m_{ch}})\right] - j \times \left[4\pi^2(f_1 - f)(f_2 - f) \times \left(\beta_{2,n_s} + \pi\beta_{3,n_s}(f_1 + f_2 - 2f_{n_s}^c)\right)\right]}$$

; for $f_{s,m_{ch}} \leq f_1 \leq f_{e,m_{ch}}$ and $f_{s,CUT} \leq f_2 \leq f_{e,CUT}$

Also, having eq. (46) we can see:

$$e^{\int_0^{L_S(p)} \kappa_p(z,f_{CUT}) dz} = e^{-j \times \int_0^{L_S(p)} \beta_p(z,f_{CUT}) dz} \times e^{-\int_0^{L_S(p)} \alpha_p(z,f_{CUT}) dz} \quad eq.\ (64)$$

And furthermore:

$$e^{\int_0^{L_S(p)}[\kappa_p(z,f_1)+\kappa_p(z,f_2)+\kappa_p^*(z,\ f_1+f_2-f_{CUT})] dz} \quad eq.\ (65)$$

$$= e^{-j \int_0^{L_S(p)} \left(\beta_p(z,f_1)+\beta_p(z,f_2)-\beta_p(z,f_3)\right)dz}$$
$$\times e^{-\int_0^{L_S(p)} \alpha_p(z,f_{c,m_{ch}}) dz} \times e^{-\int_0^{L_S(p)} \alpha_p(z,f_{CUT}) dz}$$
$$\times e^{-\int_0^{L_S(p)} \alpha_p(z,f_{c,m_{ch}}) dz}$$

; for $f_{s,m_{ch}} \leq f_1 \leq f_{e,m_{ch}}$ and $f_{s,CUT} \leq f_2 \leq f_{e,CUT}$

Therefore, the link function in (45) can be simplified using equations (63), (64) and (65) as:

$$LK(f_1, f_2, f_1 + f_2 - f_{CUT}) \cong -j \sum_{n_s=1}^{N_S} \xi_{n_s}(f_1, f_2, f_1 + f_2 - f_{CUT}) \quad eq.\ (66)$$

; for $f_{s,m_{ch}} \leq f_1 \leq f_{e,m_{ch}}$ and $f_{s,CUT} \leq f_2 \leq f_{e,CUT}$

Where in eq. (66):

$$\xi_{n_s}(f_1, f_2, f_1 + f_2 - f_{CUT}) \triangleq \gamma_{n_s} \times e^{-\frac{2\alpha_1^{(n_s)}(f_{c,m_{ch}})}{\sigma^{(n_s)}(f_{c,m_{ch}})}} \times \sum_{k=0}^{M(n_s,f_{c,m_{ch}})} \frac{1}{k!} \times \left[\frac{2\alpha_1^{(n_s)}(f_{c,m_{ch}})}{\sigma^{(n_s)}(f_{c,m_{ch}})}\right]^k \times \quad eq.\ (67)$$

$$\frac{1}{\left[2\alpha_0^{(n_s)}(f_{c,m_{ch}}) + k \times \sigma^{(n_s)}(f_{c,m_{ch}})\right] - j \times \left[4\pi^2(f_1-f)(f_2-f) \times \left(\beta_{2,n_s} + \pi\beta_{3,n_s}(f_1+f_2-2f_{n_s}^c)\right)\right]}$$

$$\left\{\prod_{p=n_s}^{N_s} \Gamma_p^{\frac{1}{2}}(f_1+f_2-f_3)\, e^{j\theta_p(f_1+f_2-f_3)} \times e^{-j\times\int_0^{L_S(p)}\beta_p(z,f_{CUT})\,dz} \times e^{-\int_0^{L_S(p)}\alpha_p(z,f_{CUT})\,dz}\, e^{-j\beta_{DCU}^{(p)}(f_1+f_2-f_3)}\right\}$$

$$\times \left\{\prod_{p=1}^{n_s-1}\left\{[\Gamma_p(f_1)\Gamma_p(f_2)\Gamma_p(f_3)]^{\frac{1}{2}} e^{j[\theta_p(f_1)+\theta_p(f_2)-\theta_p(f_3)]} \times e^{-j\int_0^{L_S(p)}\left(\beta_p(z,f_1)+\beta_p(z,f_2)-\beta_p(z,f_3)\right)dz}\right.\right.$$

$$\times e^{-\int_0^{L_S(p)}\alpha_p(z,f_{c,m_{ch}})\,dz} \times e^{-\int_0^{L_S(p)}\alpha_p(z,f_{CUT})\,dz} \times e^{-\int_0^{L_S(p)}\alpha_p(z,f_{c,m_{ch}})\,dz}$$

$$\left.\left.\times e^{-j\left[\beta_{DCU}^{(p)}(f_1)+\beta_{DCU}^{(p)}(f_2)-\beta_{DCU}^{(p)}(f_3)\right]}\right\}\right\}$$

; for $f_{s,m_{ch}} \leq f_1 \leq f_{e,m_{ch}}$ and $f_{s,CUT} \leq f_2 \leq f_{e,CUT}$

As it is obvious from eq. (44), for the calculation of the GN model, we need to calculate $|LK(f_1, f_2, f_1+f_2-f_{CUT})|^2$. Therefore, by using eq. (66):

$$|LK(f_1, f_2, f_1+f_2-f_{CUT})|^2$$
$$= \sum_{n_s=1}^{N_s}\sum_{n_s'=1}^{N_s} \xi_{n_s}(f_1,f_2,f_1+f_2-f_{CUT}) \times \xi_{n_s'}^*(f_1,f_2,f_1+f_2-f_{CUT})$$
$$= \sum_{n_s=1}^{N_s} |\xi_{n_s}(f_1,f_2,f_1+f_2-f_{CUT})|^2$$
$$+ \sum_{n_s=1}^{N_s}\sum_{\substack{n_s'=1 \\ n_s' \neq n_s}}^{N_s} \xi_{n_s}(f_1,f_2,f_1+f_2-f_{CUT}) \times \xi_{n_s'}^*(f_1,f_2,f_1+f_2-f_{CUT})$$

eq. (68)

; for $f_{s,m_{ch}} \leq f_1 \leq f_{e,m_{ch}}$ and $f_{s,CUT} \leq f_2 \leq f_{e,CUT}$

In the incoherent GN model which we use in this paper the second term at the right side of eq. (68) is ignored and:

$$|LK(f_1,f_2,f_1+f_2-f_{CUT})|^2 \cong \sum_{n_s=1}^{N_s} |\xi_{n_s}(f_1,f_2,f_1+f_2-f_{CUT})|^2$$

eq. (69)

; for $f_{s,m_{ch}} \leq f_1 \leq f_{e,m_{ch}}$ and $f_{s,CUT} \leq f_2 \leq f_{e,CUT}$

Combining equations (67) and (69):

$$|LK(f_1,f_2,f_1+f_2-f_{CUT})|^2 \cong \sum_{n_s=1}^{N_s} |\xi_{n_s}(f_1,f_2,f_1+f_2-f_{CUT})|^2$$

eq. (70)

$$= \sum_{n_s=1}^{N_s} \gamma_{n_s}^2 \times e^{-\frac{4\alpha_1^{(n_s)}(f_{c,m_{ch}})}{\sigma^{(n_s)}(f_{c,m_{ch}})}} \times |\zeta|^2 \times \{\prod_{p=n_s}^{N_s} \Gamma_p(f_{CUT}) \times S_p^{-1}(f_{CUT})\} \times$$

$$\left\{ \prod_{p=1}^{n_s-1} \Gamma_p^2(f_{m_{ch}}) \Gamma_p(f_{CUT}) \times S_p^{-2}(f_{m\_ch}) \times S_p^{-1}(f_{CUT}) \right\}$$

$$; \text{ for } f_{s,m_{ch}} \leq f_1 \leq f_{e,m_{ch}} \text{ and } f_{s,CUT} \leq f_2 \leq f_{e,CUT}$$

Where in eq. (70), $S_p(f)$ denotes total power loss at the fiber span p and frequency f which is defined as:

$$S_p(f) \triangleq \frac{P(z=0,f)}{P(z=L_s(p),f)} = e^{+2\int_0^{L_s(p)} \alpha_p(z,f)\, dz}$$

eq. (71)

Where $L_s(p)$ is the physical length due to fiber span p in the fiber link. Also $\zeta$ in eq. (7) is:

$$\zeta = \sum_{k=0}^{M(n_s, f_{c,m_{ch}})} \frac{1}{k!} \times \left[\frac{2\alpha_1^{(n_s)}(f_{c,m_{ch}})}{\sigma^{(n_s)}(f_{c,m_{ch}})}\right]^k \times \frac{1}{[2\alpha_0^{(n_s)}(f_{c,m_{ch}}) + k \times \sigma^{(n_s)}(f_{c,m_{ch}})] - j \times \varrho}$$

eq. (72)

Where $\varrho$ in (72):

$$\varrho \triangleq \left[ 4\pi^2 (f_1 - f_{CUT})(f_2 - f_{CUT}) \times \left( \beta_{2,n_s} + \pi \beta_{3,n_s}(f_1 + f_2 - 2f_{n_s}^c) \right) \right]$$

eq. (73)

Therefore $|\zeta|^2$ is calculated as:

$$|\zeta|^2 = \sum_{k_1=0}^{M(n_s, f_{c,m_{ch}})} \sum_{k_2=0}^{M(n_s, f_{c,m_{ch}})} \frac{2}{k_1! k_2!} \times \left[\frac{2\alpha_1^{(n_s)}(f_{c,m_{ch}})}{\sigma^{(n_s)}(f_{c,m_{ch}})}\right]^{k_1+k_2}$$

$$\times \frac{1}{4\alpha_0^{(n_s)}(f_{c,m_{ch}}) + (k_1+k_2)\sigma^{(n_s)}(f_{c,m_{ch}})}$$

$$\times \frac{2\alpha_0^{(n_s)}(f_{c,m_{ch}}) + k_1 \sigma^{(n_s)}(f_{c,m_{ch}})}{\left[2\alpha_0^{(n_s)}(f_{c,m_{ch}}) + k_1 \times \sigma^{(n_s)}(f_{c,m_{ch}})\right]^2 + \varrho^2}$$

eq. (74)

Eq. (74) can be further simplified as:

$$|\zeta|^2 = \sum_{k_1=0}^{M(n_s,f_{c,m_{ch}})} \frac{1}{k_1!} \times \left[\frac{2\alpha_1^{(n_s)}(f_{c,m_{ch}})}{\sigma^{(n_s)}(f_{c,m_{ch}})}\right]^{k_1} \times \frac{2\alpha_0^{(n_s)}(f_{c,m_{ch}}) + k_1\sigma^{(n_s)}(f_{c,m_{ch}})}{\left[2\alpha_0^{(n_s)}(f_{c,m_{ch}}) + k_1 \times \sigma^{(n_s)}(f_{c,m_{ch}})\right]^2 + \varrho^2} \times h(k_1, n_s, f_{m_{ch}}) \quad eq.\ (75)$$

Where $h(k_1, n_s, f_{m_{ch}})$ in eq. (75) is defined as:

$$h(k_1, n_s, f_{m_{ch}}) \triangleq \sum_{k_2=0}^{M(n_s,f_{c,m_{ch}})} \frac{2}{k_2!} \times \left[\frac{2\alpha_1^{(n_s)}(f_{c,m_{ch}})}{\sigma^{(n_s)}(f_{c,m_{ch}})}\right]^{k_2} \times \frac{1}{4\alpha_0^{(n_s)}(f_{c,m_{ch}}) + (k_1+k_2)\sigma^{(n_s)}(f_{c,m_{ch}})} \quad eq.\ (76)$$

Substituting eq. (75) in eq. (70) we have:

$$|LK(f_1, f_2, f_1 + f_2 - f_{CUT})|^2 \cong \sum_{n_s=1}^{N_s} |\xi_{n_s}(f_1, f_2, f_1 + f_2 - f_{CUT})|^2 \quad eq.\ (77)$$

$$= \sum_{n_s=1}^{N_s} \gamma_{n_s}^2 \times \left\{\prod_{p=1}^{n_s-1} \Gamma_p^2(f_{m_{ch}})\Gamma_p(f_{CUT}) \times S_p^{-2}(f_{m_{ch}}) \times S_p^{-1}(f_{CUT})\right\} \times \left\{\prod_{p=n_s}^{N_s} \Gamma_p(f_{CUT}) \times S_p^{-1}(f_{CUT})\right\}$$

$$\times e^{-\frac{4\alpha_1^{(n_s)}(f_{c,m_{ch}})}{\sigma^{(n_s)}(f_{c,m_{ch}})}} \sum_{k_1=0}^{M(n_s,f_{c,m_{ch}})} \frac{h(k_1, n_s, f_{m_{ch}})}{k_1!} \times \left[\frac{2\alpha_1^{(n_s)}(f_{c,m_{ch}})}{\sigma^{(n_s)}(f_{c,m_{ch}})}\right]^{k_1}$$

$$\times \frac{2\alpha_0^{(n_s)}(f_{c,m_{ch}}) + k_1\sigma^{(n_s)}(f_{c,m_{ch}})}{\left[2\alpha_0^{(n_s)}(f_{c,m_{ch}}) + k_1 \times \sigma^{(n_s)}(f_{c,m_{ch}})\right]^2 + \varrho^2}$$

*; for $f_{s,m_{ch}} \leq f_1 \leq f_{e,m_{ch}}$ and $f_{s,CUT} \leq f_2 \leq f_{e,CUT}$*

Substituting eq. (77) in eq. (44) we have:

$$G_{NLI}(f_{CUT}) \cong \frac{16}{27} \times$$
$$\sum_{n_s=1}^{N_s} \gamma_{n_s}^2 \times \sum_{m_{ch}=1}^{N_c} G_{CUT} G_{m_{ch}}^2 \times (2 - \delta_{CUT,m_{ch}}) \times$$
$$\left\{ \prod_{p=1}^{n_s-1} \Gamma_p^2(f_{m_{ch}}) \Gamma_p(f_{CUT}) \times S_p^{-2}(f_{m_{ch}}) \times S_p^{-1}(f_{CUT}) \right\} \times \left\{ \prod_{p=n_s}^{N_s} \Gamma_p(f_{CUT}) \times S_p^{-1}(f_{CUT}) \right\}$$
$$\times e^{-\frac{4\alpha_1^{(n_s)}(f_{c,m_{ch}})}{\sigma^{(n_s)}(f_{c,m_{ch}})}} \sum_{k_1=0}^{M(n_s,f_{c,m_{ch}})} \frac{h(k_1, n_s, f_{m_{ch}})}{k_1!} \times \left[ \frac{2\alpha_1^{(n_s)}(f_{c,m_{ch}})}{\sigma^{(n_s)}(f_{c,m_{ch}})} \right]^{k_1}$$
$$\times \int_{f_{s,CUT}}^{f_{e,CUT}} \int_{f_{s,m_{ch}}}^{f_{e,m_{ch}}} \frac{2\alpha_0^{(n_s)}(f_{c,m_{ch}}) + k_1 \sigma^{(n_s)}(f_{c,m_{ch}})}{\left[2\alpha_0^{(n_s)}(f_{c,m_{ch}}) + k_1 \times \sigma^{(n_s)}(f_{c,m_{ch}})\right]^2 + \varrho^2} df_1 df_2$$

eq. (78)

For analytically solving the double integral in eq. (78) we first define the effective dispersion parameter as:

$$\beta_{2,eff}^{(n_s)}(f_{c,m_{ch}}, f_{CUT}) \triangleq \beta_{2,n_s} + \pi\beta_{3,n_s}(f_{c,m_{ch}} + f_{CUT} - 2f_{n_s}^c) \qquad eq.\ (79)$$

For achieving double integral solution, we make the below approximation [11], [12]:

$$\beta_{2,n_s} + \pi\beta_{3,n_s}(f_1 + f_2 - 2f_{n_s}^c) \cong \beta_{2,eff}^{(n_s)}(f_{c,m_{ch}}, f_{CUT})$$
$$;\ for\ f_{s,m_{ch}} \le f_1 \le f_{e,m_{ch}}\ and\ f_{s,CUT} \le f_2 \le f_{e,CUT} \qquad eq.\ (80)$$

The above approximation is in fact replacing the frequency dependent dispersion in an integration island with the dispersion value at the center of the island.

Using the approximation made in eq. (80), the double integral in eq. (78) is expressed approximately as [23]:

$$\int_{f_{s,CUT}}^{f_{e,CUT}} \int_{f_{s,m_{ch}}}^{f_{e,m_{ch}}} \frac{2\alpha_0^{(n_s)}(f_{c,m_{ch}}) + k_1 \sigma^{(n_s)}(f_{c,m_{ch}})}{\left[2\alpha_0^{(n_s)}(f_{c,m_{ch}}) + k_1 \times \sigma^{(n_s)}(f_{c,m_{ch}})\right]^2 + \varrho^2} df_1 df_2 \cong$$
$$\int_{(f_{s,CUT}-f_{CUT})}^{(f_{e,CUT}-f_{CUT})} \int_{(f_{s,m_{ch}}-f_{CUT})}^{(f_{e,m_{ch}}-f_{CUT})} \frac{\left(2\alpha_0^{(n_s)}(f_{c,m_{ch}}) + k_1 \sigma^{(n_s)}(f_{c,m_{ch}})\right) \times df_1' df_2'}{\left[2\alpha_0^{(n_s)}(f_{c,m_{ch}}) + k_1 \times \sigma^{(n_s)}(f_{c,m_{ch}})\right]^2 + \left[4\pi^2 f_1' f_2' \times \beta_{2,eff}^{(n_s)}(f_{c,m_{ch}}, f_{CUT})\right]^2}$$

eq. (81)

$$= \frac{1}{2 \times 4\pi^2 \times \beta_{2,eff}^{(n_s)}\left(f_{c,m_{ch}}, f_{CUT}\right)}$$

$$\times \left\{ F_{int}\left( \frac{4\pi^2 \times \beta_{2,eff}^{(n_s)}\left(f_{c,m_{ch}}, f_{CUT}\right) \times \left(f_{e,CUT} - f_{CUT}\right) \times \left(f_{e,m_{ch}} - f_{CUT}\right)}{2\alpha_0^{(n_s)}\left(f_{c,m_{ch}}\right) + k_1 \times \sigma^{(n_s)}\left(f_{c,m_{ch}}\right)} \right) \right.$$

$$+ F_{int}\left( \frac{4\pi^2 \times \beta_{2,eff}^{(n_s)}\left(f_{c,m_{ch}}, f_{CUT}\right) \times \left(f_{s,CUT} - f_{CUT}\right) \times \left(f_{s,m_{ch}} - f_{CUT}\right)}{2\alpha_0^{(n_s)}\left(f_{c,m_{ch}}\right) + k_1 \times \sigma^{(n_s)}\left(f_{c,m_{ch}}\right)} \right)$$

$$- F_{int}\left( \frac{4\pi^2 \times \beta_{2,eff}^{(n_s)}\left(f_{c,m_{ch}}, f_{CUT}\right) \times \left(f_{e,CUT} - f_{CUT}\right) \times \left(f_{s,m_{ch}} - f_{CUT}\right)}{2\alpha_0^{(n_s)}\left(f_{c,m_{ch}}\right) + k_1 \times \sigma^{(n_s)}\left(f_{c,m_{ch}}\right)} \right)$$

$$\left. - F_{int}\left( \frac{4\pi^2 \times \beta_{2,eff}^{(n_s)}\left(f_{c,m_{ch}}, f_{CUT}\right) \times \left(f_{s,CUT} - f_{CUT}\right) \times \left(f_{e,m_{ch}} - f_{CUT}\right)}{2\alpha_0^{(n_s)}\left(f_{c,m_{ch}}\right) + k_1 \times \sigma^{(n_s)}\left(f_{c,m_{ch}}\right)} \right) \right\}$$

where:

$$F_{int}(x) \triangleq j \times \{Li_2(-jx) - Li_2(jx)\} \cong \pi \, \text{asinh}\left(\frac{x}{2}\right) \qquad \text{eq. (82)}$$

The special function $Li_2(.)$ is the *polylogarithm* function or *Jonquière's* function of order two [24]. Alternatively, eq. (81) can be written as:

$$\int_{f_{s,CUT}}^{f_{e,CUT}} \int_{f_{s,m_{ch}}}^{f_{e,m_{ch}}} \frac{2\alpha_0^{(n_s)}\left(f_{c,m_{ch}}\right) + k_1 \sigma^{(n_s)}\left(f_{c,m_{ch}}\right)}{\left[2\alpha_0^{(n_s)}\left(f_{c,m_{ch}}\right) + k_1 \times \sigma^{(n_s)}\left(f_{c,m_{ch}}\right)\right]^2 + \varrho^2} df_1 df_2 \qquad \text{eq. (83)}$$

$$\cong \frac{1}{2 \times 4\pi^2 \times \beta_{2,eff}^{(n_s)}\left(f_{c,m_{ch}}, f_{CUT}\right)}$$

$$\times \left\{ F_{int}\left( \frac{2\pi^2 \times \beta_{2,eff}^{(n_s)}\left(f_{c,m_{ch}}, f_{CUT}\right) \times BW_{CUT} \times \left(f_{e,m_{ch}} - f_{CUT}\right)}{2\alpha_0^{(n_s)}\left(f_{c,m_{ch}}\right) + k_1 \times \sigma^{(n_s)}\left(f_{c,m_{ch}}\right)} \right) \right.$$

$$+ F_{int}\left( \frac{2\pi^2 \times \beta_{2,eff}^{(n_s)}\left(f_{c,m_{ch}}, f_{CUT}\right) \times (-BW_{CUT}) \times \left(f_{s,m_{ch}} - f_{CUT}\right)}{2\alpha_0^{(n_s)}\left(f_{c,m_{ch}}\right) + k_1 \times \sigma^{(n_s)}\left(f_{c,m_{ch}}\right)} \right)$$

$$- F_{int}\left( \frac{2\pi^2 \times \beta_{2,eff}^{(n_s)}\left(f_{c,m_{ch}}, f_{CUT}\right) \times BW_{CUT} \times \left(f_{s,m_{ch}} - f_{CUT}\right)}{2\alpha_0^{(n_s)}\left(f_{c,m_{ch}}\right) + k_1 \times \sigma^{(n_s)}\left(f_{c,m_{ch}}\right)} \right)$$

$$\left. - F_{int}\left( \frac{2\pi^2 \times \beta_{2,eff}^{(n_s)}\left(f_{c,m_{ch}}, f_{CUT}\right) \times (-BW_{CUT}) \times \left(f_{e,m_{ch}} - f_{CUT}\right)}{2\alpha_0^{(n_s)}\left(f_{c,m_{ch}}\right) + k_1 \times \sigma^{(n_s)}\left(f_{c,m_{ch}}\right)} \right) \right\}$$

where $BW_{CUT} = f_{e,CUT} - f_{s,CUT}$ is the bandwidth (equal to Baud Rate) of the channel under test in eq. (82). Therefore, by using eq. (82), eq. (78) is written as:

$$G_{NLI}(f_{CUT}) \cong \frac{16}{27} \times$$
$$\sum_{n_s=1}^{N_s} \gamma_{n_s}^2 \times \sum_{m_{ch}=1}^{N_c} G_{CUT} G_{m_{ch}}^2 \times (2 - \delta_{CUT,m_{ch}}) \times$$
$$\left\{ \prod_{p=1}^{n_s-1} \Gamma_p^2(f_{m_{ch}}) \Gamma_p(f_{CUT}) \times S_p^{-2}(f_{m_{ch}}) \times S_p^{-1}(f_{CUT}) \right\} \times \left\{ \prod_{p=n_s}^{N_s} \Gamma_p(f_{CUT}) \times S_p^{-1}(f_{CUT}) \right\}$$
$$\times e^{-\frac{4\alpha_1^{(n_s)}(f_{c,m_{ch}})}{\sigma^{(n_s)}(f_{c,m_{ch}})}} \sum_{k_1=0}^{M(n_s,f_{c,m_{ch}})} \frac{h(k_1, n_s, f_{m_{ch}})}{k_1!} \times \left[ \frac{2\alpha_1^{(n_s)}(f_{c,m_{ch}})}{\sigma^{(n_s)}(f_{c,m_{ch}})} \right]^{k_1}$$
$$\times \frac{1}{2 \times 4\pi^2 \times \beta_{2,eff}^{(n_s)}(f_{c,m_{ch}}, f_{CUT})}$$
$$\times \left\{ F_{int}\left( \frac{2\pi^2 \times \beta_{2,eff}^{(n_s)}(f_{c,m_{ch}}, f_{CUT}) \times BW_{CUT} \times (f_{e,m_{ch}} - f_{CUT})}{2\alpha_0^{(n_s)}(f_{c,m_{ch}}) + k_1 \times \sigma^{(n_s)}(f_{c,m_{ch}})} \right) \right.$$
$$+ F_{int}\left( \frac{2\pi^2 \times \beta_{2,eff}^{(n_s)}(f_{c,m_{ch}}, f_{CUT}) \times (-BW_{CUT}) \times (f_{s,m_{ch}} - f_{CUT})}{2\alpha_0^{(n_s)}(f_{c,m_{ch}}) + k_1 \times \sigma^{(n_s)}(f_{c,m_{ch}})} \right)$$
$$- F_{int}\left( \frac{2\pi^2 \times \beta_{2,eff}^{(n_s)}(f_{c,m_{ch}}, f_{CUT}) \times BW_{CUT} \times (f_{s,m_{ch}} - f_{CUT})}{2\alpha_0^{(n_s)}(f_{c,m_{ch}}) + k_1 \times \sigma^{(n_s)}(f_{c,m_{ch}})} \right)$$
$$\left. - F_{int}\left( \frac{2\pi^2 \times \beta_{2,eff}^{(n_s)}(f_{c,m_{ch}}, f_{CUT}) \times (-BW_{CUT}) \times (f_{e,m_{ch}} - f_{CUT})}{2\alpha_0^{(n_s)}(f_{c,m_{ch}}) + k_1 \times \sigma^{(n_s)}(f_{c,m_{ch}})} \right) \right\}$$

eq. (84)

As $F_{int}(x)$ is an odd function with respect to $x$, eq. (84) can be simplified as:

$$G_{NLI}(f_{CUT}) \cong \frac{16}{27} \times$$
$$\sum_{n_s=1}^{N_s} \gamma_{n_s}^2 \times \sum_{m_{ch}=1}^{N_c} G_{CUT} G_{m_{ch}}^2 \times (2 - \delta_{CUT,m_{ch}}) \times$$
$$\left\{ \prod_{p=1}^{n_s-1} \Gamma_p^2(f_{m_{ch}}) \Gamma_p(f_{CUT}) \times S_p^{-2}(f_{m_{ch}}) \times S_p^{-1}(f_{CUT}) \right\} \times \left\{ \prod_{p=n_s}^{N_s} \Gamma_p(f_{CUT}) \times S_p^{-1}(f_{CUT}) \right\}$$

eq. (85)

$$\times e^{-\frac{4\alpha_1^{(n_s)}(f_{c,m_{ch}})}{\sigma^{(n_s)}(f_{c,m_{ch}})}} \sum_{k_1=0}^{M(n_s,f_{c,m_{ch}})} \frac{h(k_1,n_s,f_{m_{ch}})}{k_1!} \times \left[\frac{2\alpha_1^{(n_s)}(f_{c,m_{ch}})}{\sigma^{(n_s)}(f_{c,m_{ch}})}\right]^{k_1}$$

$$\times \frac{1}{4\pi^2 \times \beta_{2,eff}^{(n_s)}(f_{c,m_{ch}}, f_{CUT})}$$

$$\times \left\{ F_{int}\left(\frac{2\pi^2 \times \beta_{2,eff}^{(n_s)}(f_{c,m_{ch}}, f_{CUT}) \times BW_{CUT} \times (f_{e,m_{ch}} - f_{CUT})}{2\alpha_0^{(n_s)}(f_{c,m_{ch}}) + k_1 \times \sigma^{(n_s)}(f_{c,m_{ch}})}\right) \right.$$

$$\left. - F_{int}\left(\frac{2\pi^2 \times \beta_{2,eff}^{(n_s)}(f_{c,m_{ch}}, f_{CUT}) \times BW_{CUT} \times (f_{s,m_{ch}} - f_{CUT})}{2\alpha_0^{(n_s)}(f_{c,m_{ch}}) + k_1 \times \sigma^{(n_s)}(f_{c,m_{ch}})}\right) \right\}$$

Accepting the approximation $F_{int}(x) \cong \pi \operatorname{asinh}\left(\frac{x}{2}\right)$, eq. (85) is expressed as:

$$G_{NLI}(f_{CUT}) \cong \frac{16}{27} \times \qquad\qquad\qquad\qquad\qquad\qquad\qquad\qquad\qquad\qquad\text{eq. (86)}$$

$$\sum_{n_s=1}^{N_s} \gamma_{n_s}^2 \times \sum_{m_{ch}=1}^{N_c} G_{CUT} G_{m_{ch}}^2 \times (2 - \delta_{CUT,m_{ch}}) \times$$

$$\left\{\prod_{p=1}^{n_s-1} \Gamma_p^2(f_{m_{ch}}) \Gamma_p(f_{CUT}) \times S_p^{-2}(f_{m_{ch}}) \times S_p^{-1}(f_{CUT})\right\} \times \left\{\prod_{p=n_s}^{N_s} \Gamma_p(f_{CUT}) \times S_p^{-1}(f_{CUT})\right\}$$

$$\times e^{-\frac{4\alpha_1^{(n_s)}(f_{c,m_{ch}})}{\sigma^{(n_s)}(f_{c,m_{ch}})}} \sum_{k_1=0}^{M(n_s,f_{c,m_{ch}})} \frac{h(k_1,n_s,f_{m_{ch}})}{k_1!} \times \left[\frac{2\alpha_1^{(n_s)}(f_{c,m_{ch}})}{\sigma^{(n_s)}(f_{c,m_{ch}})}\right]^{k_1}$$

$$\times \frac{1}{4\pi \times \beta_{2,eff}^{(n_s)}(f_{c,m_{ch}}, f_{CUT})}$$

$$\times \left\{ \operatorname{asinh}\left(\frac{\pi^2 \times \beta_{2,eff}^{(n_s)}(f_{c,m_{ch}}, f_{CUT}) \times BW_{CUT} \times (f_{e,m_{ch}} - f_{CUT})}{2\alpha_0^{(n_s)}(f_{c,m_{ch}}) + k_1 \times \sigma^{(n_s)}(f_{c,m_{ch}})}\right) \right.$$

$$\left. - \operatorname{asinh}\left(\frac{\pi^2 \times \beta_{2,eff}^{(n_s)}(f_{c,m_{ch}}, f_{CUT}) \times BW_{CUT} \times (f_{s,m_{ch}} - f_{CUT})}{2\alpha_0^{(n_s)}(f_{c,m_{ch}}) + k_1 \times \sigma^{(n_s)}(f_{c,m_{ch}})}\right) \right\}$$

Another useful form of eq. (86) can be presented by considering two below equations:

$$G_{CUT}^{(n_s)} \triangleq G_{CUT} \times \{\prod_{p=1}^{n_s-1} \Gamma_p(f_{CUT}) \times S_p^{-1}(f_{CUT})\}$$

eq. (87)

$$G_{m_{ch}}^{(n_s)} \triangleq G_{m_{ch}} \times \{\prod_{p=1}^{n_s-1} \Gamma_p(f_{m_{ch}}) \times S_p^{-1}(f_{m_{ch}})\}$$

eq. (88)

Where $G_{CUT}$ and $G_{m_{ch}}$ are PSD at the launch point (start of the first fiber span) while $G_{m_{ch}}^{(n_s)}$ and $G_{CUT}^{(n_s)}$ are the PSD at the start of the $n_s$'th fiber span. Therefore, using equations (87) and (88), eq. (86) can be also written as the below form:

$$G_{NLI}(f_{CUT}) \cong \frac{16}{27} \times$$
$$\sum_{n_s=1}^{N_s} \gamma_{n_s}^2 \times \sum_{m_{ch}=1}^{N_c} G_{CUT}^{(n_s)} \times \left[G_{m_{ch}}^{(n_s)}\right]^2 \times (2 - \delta_{CUT,m_{ch}}) \times$$
$$\times \left\{\prod_{p=n_s}^{N_s} \Gamma_p(f_{CUT}) \times S_p^{-1}(f_{CUT})\right\}$$
$$\times e^{-\frac{4\alpha_1^{(n_s)}(f_{c,m_{ch}})}{\sigma^{(n_s)}(f_{c,m_{ch}})}} \sum_{k_1=0}^{M(n_s,f_{c,m_{ch}})} \frac{h(k_1,n_s,f_{m_{ch}})}{k_1!} \times \left[\frac{2\alpha_1^{(n_s)}(f_{c,m_{ch}})}{\sigma^{(n_s)}(f_{c,m_{ch}})}\right]^{k_1}$$
$$\times \frac{1}{4\pi \times \beta_{2,eff}^{(n_s)}(f_{c,m_{ch}}, f_{CUT})}$$
$$\times \left\{ asinh\left(\frac{\pi^2 \times \beta_{2,eff}^{(n_s)}(f_{c,m_{ch}}, f_{CUT}) \times BW_{CUT} \times (f_{e,m_{ch}} - f_{CUT})}{2\alpha_0^{(n_s)}(f_{c,m_{ch}}) + k_1 \times \sigma^{(n_s)}(f_{c,m_{ch}})}\right) \right.$$
$$\left. - asinh\left(\frac{\pi^2 \times \beta_{2,eff}^{(n_s)}(f_{c,m_{ch}}, f_{CUT}) \times BW_{CUT} \times (f_{s,m_{ch}} - f_{CUT})}{2\alpha_0^{(n_s)}(f_{c,m_{ch}}) + k_1 \times \sigma^{(n_s)}(f_{c,m_{ch}})}\right)\right\}$$

eq. (89)

For Comparing eq. (89) with the formula derived in [11], we may note that by making a further assumption $\left|\frac{2\alpha_1^{(n_s)}(f_{c,m_{ch}})}{\sigma^{(n_s)}(f_{c,m_{ch}})}\right| \ll 1$ for all frequencies (all WDM channels) and all fiber spans and by using eq. (58), then we would have $M(n_s, f) = 1$. If so, eq. (76) gives:

$$h(k_1, n_s, f_{m_{ch}}) = \frac{2}{4\alpha_0^{(n_s)}\left(f_{c,m_{ch}}\right) + k_1 \times \sigma^{(n_s)}\left(f_{c,m_{ch}}\right)} \quad \text{eq. (90)}$$
$$+ \frac{2}{4\alpha_0^{(n_s)}\left(f_{c,m_{ch}}\right) + (k_1 + 1) \times \sigma^{(n_s)}\left(f_{c,m_{ch}}\right)} \times \left[\frac{2\alpha_1^{(n_s)}(f_{c,m_{ch}})}{\sigma^{(n_s)}(f_{c,m_{ch}})}\right]$$

$$; for \left|\frac{2\alpha_1^{(n_s)}\left(f_{c,m_{ch}}\right)}{\sigma^{(n_s)}\left(f_{c,m_{ch}}\right)}\right| \ll 1, for\ M = 1\ and\ for\ k_1 = 0,1$$

Substituting eq. (90) in eq. (89) and replacing $e^{-\frac{4\alpha_1^{(n_s)}(f_{c,m_{ch}})}{\sigma^{(n_s)}(f_{c,m_{ch}})}} \cong \left[1 - \frac{4\alpha_1^{(n_s)}(f_{c,m_{ch}})}{\sigma^{(n_s)}(f_{c,m_{ch}})}\right]$ due to the assumption of $\left|\frac{2\alpha_1^{(n_s)}(f_{c,m_{ch}})}{\sigma^{(n_s)}(f_{c,m_{ch}})}\right| \ll 1$, we will have:

$$G_{NLI}(f_{CUT}) \cong \frac{16}{27} \times \quad \text{eq. (91)}$$
$$\sum_{n_s=1}^{N_s} \gamma_{n_s}^2 \times \sum_{m_{ch}=1}^{N_c} G_{CUT}^{(n_s)} \times \left[G_{m_{ch}}^{(n_s)}\right]^2 \times (2 - \delta_{CUT,m_{ch}}) \times$$
$$\left\{\prod_{p=n_s}^{N_s} \Gamma_p(f_{CUT}) \times S_p^{-1}(f_{CUT})\right\} \times \left[1 - \frac{4\alpha_1^{(n_s)}(f_{c,m_{ch}})}{\sigma^{(n_s)}(f_{c,m_{ch}})}\right]$$
$$\times \left\{\left[\frac{2}{4\alpha_0^{(n_s)}\left(f_{c,m_{ch}}\right)} + \frac{2}{4\alpha_0^{(n_s)}\left(f_{c,m_{ch}}\right) + \sigma^{(n_s)}\left(f_{c,m_{ch}}\right)} \times \left[\frac{2\alpha_1^{(n_s)}(f_{c,m_{ch}})}{\sigma^{(n_s)}(f_{c,m_{ch}})}\right]\right]\right.$$
$$\times \left[\frac{1}{4\pi \times \beta_{2,eff}^{(n_s)}\left(f_{c,m_{ch}}, f_{CUT}\right)}\right]$$
$$\times \left\{asinh\left(\frac{\pi^2 \times \beta_{2,eff}^{(n_s)}\left(f_{c,m_{ch}}, f_{CUT}\right) \times BW_{CUT} \times \left(f_{e,m_{ch}} - f_{CUT}\right)}{2\alpha_0^{(n_s)}\left(f_{c,m_{ch}}\right)}\right)\right.$$
$$\left.\left. - asinh\left(\frac{\pi^2 \times \beta_{2,eff}^{(n_s)}\left(f_{c,m_{ch}}, f_{CUT}\right) \times BW_{CUT} \times \left(f_{s,m_{ch}} - f_{CUT}\right)}{2\alpha_0^{(n_s)}\left(f_{c,m_{ch}}\right)}\right)\right\}\right\}$$

$$+ \left[ \frac{2}{4\alpha_0^{(n_s)}(f_{c,m_{ch}}) + \sigma^{(n_s)}(f_{c,m_{ch}})} + \frac{2}{4\alpha_0^{(n_s)}(f_{c,m_{ch}}) + 2\times\sigma^{(n_s)}(f_{c,m_{ch}})} \times \left[\frac{2\alpha_1^{(n_s)}(f_{c,m_{ch}})}{\sigma^{(n_s)}(f_{c,m_{ch}})}\right]\right]$$

$$\times \left[\frac{2\alpha_1^{(n_s)}(f_{c,m_{ch}})}{\sigma^{(n_s)}(f_{c,m_{ch}})}\right] \times \left[\frac{1}{4\pi \times \beta_{2,eff}^{(n_s)}(f_{c,m_{ch}}, f_{CUT})}\right]$$

$$\times \left\{ asinh\left(\frac{\pi^2 \times \beta_{2,eff}^{(n_s)}(f_{c,m_{ch}}, f_{CUT}) \times BW_{CUT} \times (f_{e,m_{ch}} - f_{CUT})}{2\alpha_0^{(n_s)}(f_{c,m_{ch}}) + \sigma^{(n_s)}(f_{c,m_{ch}})}\right) \right.$$

$$\left. - asinh\left(\frac{\pi^2 \times \beta_{2,eff}^{(n_s)}(f_{c,m_{ch}}, f_{CUT}) \times BW_{CUT} \times (f_{s,m_{ch}} - f_{CUT})}{2\alpha_0^{(n_s)}(f_{c,m_{ch}}) + \sigma^{(n_s)}(f_{c,m_{ch}})}\right)\right\}\right\}$$

$$; for \left|\frac{2\alpha_1^{(n_s)}(f_{c,m_{ch}})}{\sigma^{(n_s)}(f_{c,m_{ch}})}\right| \ll 1$$

Eq. (91) is the same exact formula which has been derived in [11, equations (33)-(37)]. In other words, in [11], a CFM for NLI assessment was derived based on the assumption $\left|\frac{2\alpha_1^{(n_s)}(f_{c,m_{ch}})}{\sigma^{(n_s)}(f_{c,m_{ch}})}\right| \ll 1$ for all spans and all frequencies while, in this paper, we have been able to relax the limitation $\left|\frac{2\alpha_1^{(n_s)}(f_{c,m_{ch}})}{\sigma^{(n_s)}(f_{c,m_{ch}})}\right| \ll 1$ considered in [11], and the derived formula in this paper (eq. (86) or eq. (89)) is a general formula without any limitation on the $\left|\frac{2\alpha_1^{(n_s)}(f_{c,m_{ch}})}{\sigma^{(n_s)}(f_{c,m_{ch}})}\right|$ value. This makes the formula capable of accurate modeling of WDM systems with relatively higher launched powers and for systems with co-propagating pump Raman amplification as well.

### 6- Correction of the closed form formula

Based on the approach in [13], [14], [15], [16] we can add an approximate term which considers the coherent NLI accumulation effect for the channel under test derived in [17] and two correction factors ($\rho_{CUT}^{(n_s)}, \rho_{m_{ch}}^{(n_s)}$) which make the closed-form model substantially more accurate. Eq. (89) will be corrected to:

$$G_{NLI}(f_{CUT}) \cong \frac{16}{27} \times$$
$$\sum_{n_s=1}^{N_s} \gamma_{n_s}^2 \times \sum_{m_{ch}=1}^{N_c} G_{CUT}^{(n_s)} \times \left[G_{m_{ch}}^{(n_s)}\right]^2 \times (2 - \delta_{CUT,m_{ch}})$$
$$\times \left(\rho_{CUT}^{(n_s)} \times \delta_{CUT,m_{ch}} + \rho_{m_{ch}}^{(n_s)} \times (1 - \delta_{CUT,m_{ch}})\right) \times$$

eq. (92)

$$\left\{ \prod_{p=n_s}^{N_s} \Gamma_p(f_{CUT}) \times S_p^{-1}(f_{CUT}) \right\}$$

$$\times e^{-\frac{4\alpha_1^{(n_s)}(f_{c,m_{ch}})\, M(n_s,f_{c,m_{ch}})}{\sigma^{(n_s)}(f_{c,m_{ch}})}} \sum_{k_1=0} \frac{h(k_1,n_s,f_{m_{ch}})}{k_1!} \times \left[\frac{2\alpha_1^{(n_s)}(f_{c,m_{ch}})}{\sigma^{(n_s)}(f_{c,m_{ch}})}\right]^{k_1}$$

$$\times \frac{1}{4\pi \times \beta_{2,eff}^{(n_s)}(f_{c,m_{ch}}, f_{CUT})}$$

$$\times \left\{ \text{asinh}\left( \frac{\pi^2 \times \beta_{2,eff}^{(n_s)}(f_{c,m_{ch}}, f_{CUT}) \times BW_{CUT} \times (f_{e,m_{ch}} - f_{CUT})}{2\alpha_0^{(n_s)}(f_{c,m_{ch}}) + k_1 \times \sigma^{(n_s)}(f_{c,m_{ch}})} \right) \right.$$

$$\left. - \text{asinh}\left( \frac{\pi^2 \times \beta_{2,eff}^{(n_s)}(f_{c,m_{ch}}, f_{CUT}) \times BW_{CUT} \times (f_{s,m_{ch}} - f_{CUT})}{2\alpha_0^{(n_s)}(f_{c,m_{ch}}) + k_1 \times \sigma^{(n_s)}(f_{c,m_{ch}})} \right) \right\} +$$

$$\frac{16}{27} \times \rho_{coh} \times \sum_{n_s=1}^{N_s} \gamma_{n_s}^2 \times \left[G_{CUT}^{(n_s)}\right]^3 \times \rho_{CUT}^{(n_s)} \times \left\{ \prod_{p=n_s}^{N_s} \Gamma_p(f_{CUT}) \times S_p^{-1}(f_{CUT}) \right\}$$

$$\times \frac{1}{4\pi \times \beta_{2,eff}^{(n_s)}(f_{CUT}, f_{CUT}) \times \alpha_0^{(n_s)}(f_{CUT})} \times 2$$

$$\times \frac{Si(\pi^2 \times \beta_{2,eff}^{(n_s)}(f_{CUT}, f_{CUT}) \times L_s^{(n_s)} \times BW_{CUT}^2)}{\pi \times \alpha_0^{(n_s)}(f_{CUT}) \times L_s^{(n_s)}} \times \left\{ HN(N_s - 1) + \frac{1 - N_s}{N_s} \right\}$$

The yellow highlighted term in eq. (92) is the approximate coherence term [17]. Also, $Si(.)$ is the sine integral function and $HN(.)$ is the harmonic number series. It is obvious that by setting $\rho_{co} = 0$ (ignoring the approximated coherence term) and $\rho_{CUT}^{(n_s)} = \rho_{m_{ch}}^{(n_s)} = 1$ (not imposing correction factors) eq.(92) becomes the same as eq. (89).

The factors ($\rho_{CUT}^{(n_s)}, \rho_{m_{ch}}^{(n_s)}$), reported in detail in [16], are simple laws that include a few physical quantities (accumulated dispersion, CUT symbol rate, EGN-model modulation format constant $\Phi$ [4] of each channel) and a number of free parameters whose tuning relied on machine-learning. See [16] for a detailed discussion of the approach and results.

With these modifications, we get a CFM (which we call CFM5) whose accuracy is on the order of EGN model instead of the GN model, over a wide envelope of possible WDM ultra-broadband systems.

# 7- An algorithm for NLI evaluation using the derived closed-form model CFM5

In this section, a step by step method is presented for the numerical evaluation of NLI in ultrawideband WDM systems. This algorithm is the logical result of the steps performed in the mathematical derivation of equations (84)-(87). We believe this summarized stepwise algorithm could make the software numerical implementation simpler and clearer. The main steps are shown in figure (3).

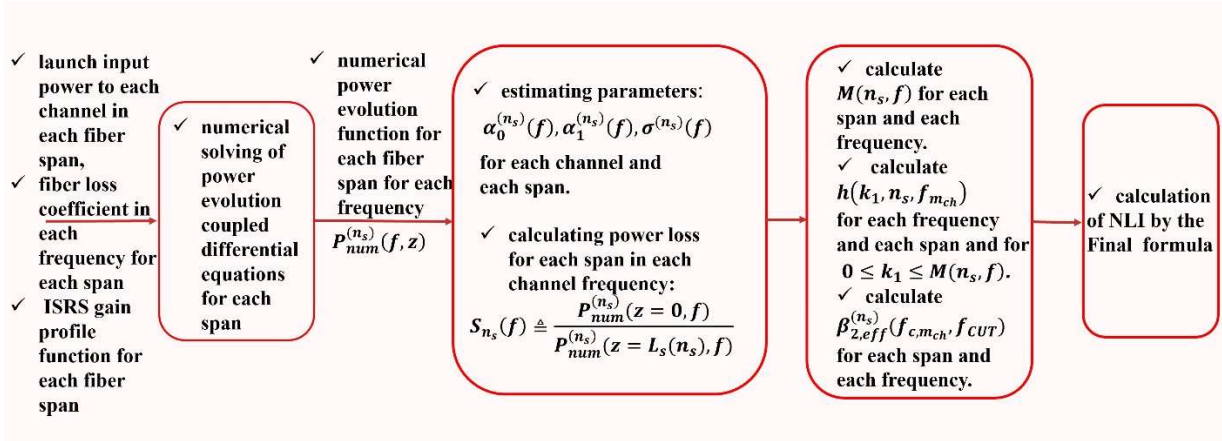

*Figure. (3): Main steps in numerical calculation of NLI*

We assume that the link that should be analyzed contains $N_s$ fiber spans and the WDM scheme Contains $N_c$ channels with center frequencies $f_{c,1}, f_{c,2}, ..., f_{c,N_c}$. The steps on detail are as:

1. We feed eq. (1) by input launch power to each span and for each channel (frequency) and also the Raman gain profile for each fiber span. Then using set of differential equations in eq. (1), we calculate the power evolution function along each span and for each channel. This can be done using famous differential equations numerical solving methods like *Runge-Kutta* [25]. Therefore we will have $P_{num}^{(n_s)}(f,z), \forall z$ for $f \epsilon \{f_{c,1}, f_{c,2}, ..., f_{c,N_c}\}$ and $n_s \epsilon \{1,2,...,N_s\}$.

2. In this step, we find power loss for each span and in each frequency using eq. (71) and therefore we have $S_{n_s}(f) \triangleq \frac{P_{num}^{(n_s)}(z=0,f)}{P_{num}^{(n_s)}(z=L_s(n_s),f)}$ for $f \epsilon \{f_{c,1}, f_{c,2}, ..., f_{c,N_c}\}$ and $n_s \epsilon \{1,2,...,N_s\}$.

3. We find $G_{m_{ch}}^{(n_s)}$ and $G_{CUT}^{(n_s)}$ for all WDM channels and all fiber spans using equations (87) and (88).

4. In this step we find the optimum $\alpha_0^{(n_s)}(f), \alpha_1^{(n_s)}(f), \sigma^{(n_s)}(f)$ for each span and each channel. For each span and each channel we first assume $\sigma^{(n_s)}(f)$ equal to an initial fixed value, find $\alpha_0^{(n_s)}(f), \alpha_1^{(n_s)}(f)$ by eq. (30.2) and then calculate cost function in eq. (24).

Using a minimization search method, like golden section search method, we search for a better $\sigma^{(n_s)}(f)$. We can search for the optimum value of $\sigma^{(n_s)}(f)$ in a reasonable numeric interval like $\sigma^{(n_s)}(f) \epsilon [\alpha^{(n_s)}(f), 4 \times \alpha^{(n_s)}(f)]$ where $\alpha^{(n_s)}(f)$ is the fiber intrinsic loss for span $n_s$ at frequency $f$ in the absence of Raman effect (In the absence of Raman effect, the z-dependent power in span $n_s$ in frequency $f$ propagates along fiber span as $P_{without-Raman}^{(n_s)}(z,f) = P^{(n_s)}(z=0,f) \times e^{-2\times\alpha^{(n_s)}(f)\times z}$). The better (more optimum) $\sigma^{(n_s)}(f)$ is the one when chosen, after calculating $\alpha_0^{(n_s)}(f), \alpha_1^{(n_s)}(f)$ by eq. (30.2) for it, the cost function in eq. (24) has lower value. So, we first make a selection for $m_c \geq 0$ in eq. (24). Then select $\sigma^{(n_s)}(f)$ in the search interval, calculate $\alpha_0^{(n_s)}(f), \alpha_1^{(n_s)}(f)$ by eq. (30.2), calculate cost function by eq. (24) and compare the value of cost function to the previous iteration. With this iterative search we can find optimum values of $\sigma^{(n_s)}(f)$, $\alpha_0^{(n_s)}(f), \alpha_1^{(n_s)}(f)$ for all channels and all spans.

5- We calculate $M(n_s, f)$ for all channels and all spans by eq. (58).
6- We calculate $h(k_1, n_s, f_{m_{ch}})$ using eq. (76) for each span and each frequency and for all values of $k_1$ in range $0 \leq k_1 \leq M(n_s, f)$.
7- We calculate effective dispersion parameter $\beta_{2,eff}^{(n_s)}(f_{c,m_{ch}}, f_{CUT})$ for all fiber spans and all WDM channels by eq. (79).
8- Using eq. (92) and the issues calculated in previous steps, we can calculate PSD of the NLI in the center frequency of channel under test (CUT).

## 9- Limitations in using the derived closed form model CFM5

The derivation of eq. (86) started from the general GN model formula in eq. (31) and along the derivation we made several assumptions and approximations. These assumptions and approximations are valid under specific conditions which may limit the application of the final derived closed form model. For having a clear view of the constraints under which we can use equation (86) reliably, we review the approximations and assumptions here:

1- As we did the double integration in the GN model for the SCI-XCI islands only and disregarded all other islands, the result is accurate when dispersion is greater than about 2 ps/(nm km) (see [16] for a more detailed discussion and error assessment results).
2- For going from eq. (61) to eq. (62) we assumed that $e^{L_s(n_s)\times\{-2\alpha_0^{(n_s)}(f_{c,m_{ch}})-k\times\sigma^{(n_s)}(f_{c,m_{ch}})\}} \ll 1$ which implies that each fiber span loss (for all frequencies in WDM comb and in the absence of Raman effect) must be large enough (greater than 8-10 dB). Therefore, for short and very short spans with smaller loss, some noticeable NLI estimation error may be present.
3- In writing eq. (69) the incoherent approximation for the GN model is used. Even though an extra approximate analytical term has been added to account for coherent NLI accumulation, certain situations cannot be dealt with, such as ultra-low dispersion or zero-dispersion propagation and dispersion-managed systems.

4- The 1st step in the calculation of NLI by the derived closed form model is finding the power evolution by numerically solving the set of differential equations in eq. (1). This must be done accurately otherwise the overall accuracy of the CFM is reduced.

## 10-   Conclusion

In this paper we have shown how to improve the closed-form model (CFM) initially proposed in [11] so that it can better handle C+L (or even broader-bandwidth) systems where ISRS is strong and possibly forward Raman amplification is present. The model, that we call CFM5, also incorporates a series of accuracy improvements that we proposed in [16].

A distinct feature of CFM5 is that its complexity can be adapted to the strength of the ISRS or forward-Raman amplification, simply by adding more terms in a summation, as needed. In any case the speed of evaluation of the NLI is essentially real-time.

What remains borderline real-time is the assessment of the channels power profile, that currently relies on the numerical integration of the forward SRS differential equation. Work is ongoing to try to speed up this phase too, to obtain a fully real-time overall model.

## 11-   Acknowledgements

This work was supported by Cisco Systems through OPTSYS 2020 contract with Politecnico di Torino and by the PhotoNext Center of Politecnico di Torino. The Authors would like to thank Stefano Piciaccia and Fabrizio Forghieri from CISCO Photonics for the fruitful discussions and interactions.## 12-   References

[1] A. Mecozzi and R.-J. Essiambre, 'Nonlinear Shannon limit in pseudolinear coherent systems,' *J. Lightwave Technol.,* vol. 30, no. 12, pp. 2011-2024, June 15th 2012.

[2] R. Dar, M. Feder, A. Mecozzi, and M. Shtaif, 'Properties of nonlinear noise in long, dispersion-uncompensated fiber links,'  *Optics Express,* vol. 21, no. 22, pp. 25685--25699, Nov. 2013.

[3] P. Poggiolini, G. Bosco, A. Carena, V. Curri,  Y. Jiang, F. Forghieri, 'The GN model of  fiber non-linear propagation and its applications,' *J. of Lightwave Technol.,* vol. 32, no. 4, pp. 694--721, Feb. 2014.

[4] A. Carena, G. Bosco, V. Curri, Y. Jiang, P. Poggiolini and F. Forghieri, 'EGN model of non-linear fiber propagation,' *Optics Express,* vol. 22, no. 13, pp. 16335-16362, June 2014. Extended appendices with full formulas derivations can be found in the version of this paper available on www.arXiv.org.

[5] P. Serena, A. Bononi, 'A Time-Domain Extended Gaussian Noise Model,*' J. Lightwave Technol.,* vol.33, no. 7, pp. 1459-1472, Apr. 2015.

[6] A. Bononi, P. Serena, N. Rossi, E. Grellier, F. Vacondio, 'Modeling nonlinearity in coherent transmissions with dominant intrachannel-four-wave-mixing,' *Optics Express,* vol. 20, pp. 7777-7791, 26 March 2012.35

[7] M. Secondini and E. Forestieri, 'Analytical fiber-optic channel model in the presence of cross-phase modulations,' *IEEE Photon. Technol. Lett.,* vol. 24, no. 22, pp. 2016-2019, Nov. 15th 2012.

[8] P. Johannisson, M. Karlsson, 'Perturbation analysis of nonlinear propagation in a strongly dispersive optical communication system,' *J. Lightwave Technol.,* vol. 31, no. 8, pp. 1273-1282, Apr. 15, 2013.

[9] R. Dar, M. Feder, A. Mecozzi, M. Shtaif, 'Pulse collision picture of inter-channel nonlinear interference noise in fiber-optic communications,' *J. Lightwave Technol.,* vol. 34, no. 2, pp. 593-607, Jan. 2016.

[10] P. Poggiolini, Y. Jiang, 'Recent Advances in the Modeling of the Impact of Nonlinear Fiber Propagation Effects on Uncompensated Coherent Transmission Systems,' tutorial review, *J. of Lightwave Technol.,* vol. 35, no. 3, pp. 458-480, Feb. 2017.

[12] D. Semrau, R. I. Killey, P. Bayvel, 'A Closed-Form Approximation of the Gaussian Noise Model in the Presence of Inter-Channel Stimulated Raman Scattering,' *www.arXiv.org,* paper arXiv:1808.07940, Aug. 23rd 2018.

[11] P. Poggiolini, 'A generalized GN-model closed-form formula,' *www.arXiv.org,* paper arXiv:1810.06545v2, Sept. 24th 2018.

[13] P. Poggiolini, M. Ranjbar Zefreh, G. Bosco, F. Forghieri, and S. Piciaccia, 'Accurate non-linearity fully-closed-form formula based on the GN/EGN model and large-data-set fitting.' *in Proc. of Optical Fiber Communication Conference (OFC),* paper M1I-4., San Diego, Mar. 2019.

[14] M. Ranjbar Zefreh, A. Carena, F. Forghieri, S. Piciaccia, P. Poggiolini, 'Accurate Closed-Form GN/EGN-model Formula Leveraging a Large QAM-System Test-Set,' *IEEE Photon. Technol. Lett.,* vol. 31, no. 16, pp. 1381-1384, Aug. 15th 2019.

[15] M. Ranjbar Zefreh, A. Carena, F. Forghieri, S. Piciaccia, P. Poggiolini, 'A GN/EGN-model real-time closed-form formula tested over 7,000 virtual links,' *Proc. of the European Conference on Optical Communications (ECOC),* paper W.1.D.3, Dublin (IR), Sept. 2019.

[16] M. Ranjbar Zefreh, F. Forghieri, S. Piciacca, P. Poggiolini, 'Accurate Closed-Form Real-Time EGN Model Formula Leveraging Machine-Learning over 8500 Thoroughly Randomized Full C-Band Systems,' *J. of Lightwave Technol.,* accepted for publication, May 2020, DOI: *10.1109/JLT.2020.2997395* .

[17] P. Poggiolini 'A Closed-Form GN-Model Non-Linear Interference Coherence Terme,' *www.arXiv.org,* paper arXiv:1906.03883, Jun. 10th 2019.

[18] Ranjbar Zefreh, Mahdi, and Pierluigi Poggiolini. "A GN-model closed-form formula considering coherency terms in the Link function and covering all possible islands in 2-D GN integration." arXiv preprint arXiv:1907.09457 (2019).

[19] Ranjbar Zefreh, Mahdi, and Pierluigi Poggiolini. "A closed-form approximate incoherent GN-model supporting MCI contributions." arXiv (2019): arXiv-1911.

[20] Tariq, Salim, and Joseph C. Palais, 'A computer model of non-dispersion-limited stimulated Raman scattering in optical fiber multiple-channel communications,' *J. of Lightwave Technology,* vol. 11, no. 12, pp. 1914-1924, 1993.

[21] Christodoulides, D. N., and R. B. Jander, 'Evolution of stimulated Raman crosstalk in wavelength division multiplexed systems,' *IEEE Photonics Technology Letters,* vol. 8, no. 12 (1996): 1722-1724.

[22] Alberto Bononi, Ronen Dar, Marco Secondini, Paolo Serena and Pierluigi Poggiolini, 'Fiber Nonlinearity and Optical System Performance,' Chapter 9 of the book: *Springer handbook of Optical Networks,* editors Biswanath Mukherjee, Ioannis Tomkos, Massimo Tornatore, Peter Winzer, Yongli Zhao, ed. Springer, 2019. ISBN 978-3-030-16250-4.

[23] Wolfram Research, Inc., Mathematica, Version 12.1, Champaign, IL (2020). URL: https://www.wolfram.com/mathematica

[24] Weisstein, Eric W. "Polylogarithm." From MathWorld – A Wolfram Web Resource. https://mathworld.wolfram.com/Polylogarithm.html

[25] Dormand, John R., and Peter J. Prince, 'A family of embedded Runge-Kutta formulae,' *Journal of computational and applied mathematics*, 6.1 (1980): 19-26.